\documentclass[pre,aps,amsmath,amssymb,amsfonts,floatfix,superscriptaddress]{revtex4} 
 
\usepackage{graphicx}
\usepackage{floatrow}
 \usepackage{color} 
 
 \definecolor{vinous}{rgb}{0.74,0.08,0.12}

\begin{document} 
 
\title{Kinetic Monte Carlo and hydrodynamic modelling of droplet dynamics on surfaces, including evaporation and condensation} 
 
\author{Mounirah Areshi} 
\affiliation{Department of Mathematical Sciences, Loughborough University, Loughborough LE11 3TU, United Kingdom} 
\affiliation{Department of Mathematical Sciences, University of Tabuk, Tabuk, Kingdom of Saudi Arabia} 

\author{Dmitri Tseluiko}
\affiliation{Department of Mathematical Sciences, Loughborough University, Loughborough LE11 3TU, United Kingdom}
\affiliation{Interdisciplinary Centre for Mathematical Modelling, Loughborough University, Loughborough LE11 3TU, United Kingdom} 
 
\author{Andrew J. Archer} 
\email{A.J.Archer@lboro.ac.uk} 
\affiliation{Department of Mathematical Sciences, Loughborough University, Loughborough LE11 3TU, United Kingdom}
\affiliation{Interdisciplinary Centre for Mathematical Modelling, Loughborough University, Loughborough LE11 3TU, United Kingdom}
 
\begin{abstract} 
We present a lattice-gas (generalised Ising) model for liquid droplets on solid surfaces. The time evolution in the model involves two processes: (i) Single-particle moves which are determined by a kinetic Monte Carlo algorithm. These incorporate into the model particle diffusion over the surface and within the droplets and also evaporation and condensation, i.e.\ the exchange of particles between droplets and the surrounding vapour. (ii) Larger-scale collective moves, modelling advective hydrodynamic fluid motion, determined by considering the dynamics predicted by a thin-film equation. The model enables us to relate how macroscopic quantities such as the contact angle and the surface tension depend on the microscopic interaction parameters between the particles and with the solid surface. We present results for droplets joining, spreading, sliding under gravity, dewetting, the effects of evaporation, the interplay of diffusive and advective dynamics, and how all this behaviour depends on the temperature and other parameters.
\end{abstract} 
 
 
\maketitle 
  \section{Introduction}\label{sec:1}

Understanding and predicting the collective dynamics of fluid particles (atoms or molecules) over substrates is a crucial aspect of surface science. For instance, the rate of diffusion of the particles is a key controlling factor in many dynamical processes occurring on surfaces, such as spreading, chemical reactions and the growth of islands and epitaxial layers \cite{ala2002collective}. Factors such as the temperature, the shape of the particles, the degree to which they adhere to the surface, how they adhere to each other and the particle density on the surface, greatly influence the rate of diffusion over the surface. Whether the motion over the surface is largely a single-particle phenomenon or a collective process is fundamental to determining the structures that are formed on surfaces.

The situations of particular interest here concern the behaviour of liquid droplets on solid surfaces and the influence of evaporation and wettability. This is a common situation arising in a vast range of applications, such as in coating, printing and lubrication, making such systems important to study and understand \cite{de2013capillarity}. During the drying process droplets can interact, joining and fusing in a coarsening process that involves hard to observe phenomena such as the evaporation and condensation of particles and the diffusion of particles over the surface. This behaviour is even more rich if the liquid has a tendency to dewet from the substrate. To describe situations like this, here we develop a coupled kinetic Monte Carlo (KMC) and hydrodynamic model for liquids on solid surfaces that incorporates the thermodynamics of evaporation and condensation, as well as the hydrodynamics of fluid flow. We assume a simple lattice-gas (generalised Ising) model for the underlying Hamiltonian that incorporates the microscopic interactions between the fluid particles and the surface. The benefit of modelling in this way is that although it is a simplified (coarse grained) model, it does correctly include much of the relevant fluid structure and thermodynamics and allows for the study of the interfacial properties of the model system. For example, it is fairly straightforward to find the wetting or dewetting behaviour of the liquid and the connection to the microscopic  properties of the system such as the particle interaction energies.

The use of lattice models to study equilibrium fluid wetting behaviour goes back to the 1980s \cite{pandit1982surfaces, pandit1982systematics, binder1988wetting} and they have been extremely useful for building our understanding of the physics. More recently, dynamical lattice models for non-equilibrium droplets and liquid films on surfaces have been developed \cite{rabani2003, sztrum2005, pauliac2008, vancea08, blunt2010patterns, stannard2011dewetting, vancea2011pattern, kim2011, jung2014, crivoi2014, tewes2017comparing, chalmers2017modelling}, with the focus largely being on modelling and understanding the patterns formed by the drying of particle suspensions, i.e.\ the coffee ring stain effect, etc. Some of these models solely implement a dynamics governed by the Metropolis Monte Carlo algorithm (of course, based on the model Hamiltonian), whilst other models add other kinds of particle moves, with the aim being to incorporate the influence of the fluid hydrodynamics, which leads to particle advection. For example, the model in \cite{kim2011} biases particle moves depending on the distance to the nearest contact line and the model in \cite{crivoi2014} assumes a certain form for the fluid velocity field and uses this to bias particle moves. Such moves do not satisfy detailed balance, which is required if one is to use Monte Carlo methods to simulate the dynamic properties of Hamiltonian systems \cite{fichthorn1991theoretical}. However, it is not clear to us that one should enforce detailed balance when constructing coarse grained models of such \emph{non-equilibrium} systems, since droplets on surfaces are highly dissipative systems, due to the evaporation, the strong coupling to the substrate, etc. Of course, at \emph{equilibrium}, there should be detailed balance in the model.

To model collective hydrodynamic motion in our model, we generate large-scale collective particle moves by considering the dynamics that is predicted by a thin-film equation. This is a time evolution equation for the liquid film thickness profile $h$, 
i.e. the height of the liquid-vapour interface above the surface  \cite{de2013capillarity, oron1997long, kalliadasis2007, thiele2007structure, craster2009dynamics}. This equation is derived from the Navier-Stokes equations for a film of liquid on a surface by making a long-wave approximation, which greatly simplifies the analysis. When generating these `thin-film moves', we first determine the liquid film height profile from the configuration of occupied lattice sites, we then evolve the thin-film equation for a short amount of time, and then map back to the lattice. In our model, an important quantity is the parameter $\lambda$, defined below in Eq.~\eqref{eq:lambda}, which is proportional to the ratio of the number of KMC attempted particle moves to the number of thin-film moves. When $\lambda=0$, our model reduces to just evolving the thin-film equation, whilst in the limit $\lambda\to\infty$, our model is a pure KMC model. Therefore, this parameter $\lambda\propto D/\eta$, where $D$ is a single-particle diffusion coefficient and $\eta$ is the viscosity.

Key quantities that characterises how a liquid wets a surface are the spreading parameter $S=\gamma_{sv}-(\gamma_{sl}+\gamma_{lv})$, where $\gamma_{sv}$, $\gamma_{sl}$, and $\gamma_{lv}$ are the solid-vapour, solid-liquid and liquid-vapour interfacial tensions, respectively \cite{de2013capillarity}, and also the contact angle $\theta$, which is given by Young's equation \cite{de2013capillarity}:
\begin{equation}
\label{eq:Young}
\gamma_{lv}\cos{\theta}=\gamma_{sv}-\gamma_{sl}.
\end{equation}
The contact angle is the angle that the height profile for equilibrium droplets makes with the surface. $\gamma_{lv}$ is a parameter that must be input into the thin-film equation, whilst $\theta$ is determined by the disjoining pressure $\Pi(h)$, which is a function that must also be input into the thin-film equation. All these quantities depend on the fluid state point (i.e.\ the temperature and the pressure) and also on the nature of the interactions between the fluid particles and with the substrate. Here, we use the approach of Refs.~\cite{archer2011nucleation, hughes2015liquid} (see also \cite{hughes2017, buller2017nudged}) that uses classical density functional theory (DFT) \cite{evans1979nature, evans92, hansen2013theory} to calculate the binding potential $g(h)$, from which one obtains $\Pi(h)=-\partial g(h)/\partial h$, and also the interfacial tensions. Thin-film equations together with binding potentials from DFT have been used previously to elucidated the influence of the fluid microscopic structure on droplet spreading \cite{yin2017films}. Here, we instead use the thin-film equation with the quantities $g(h)$ and $\gamma_{lv}$ determined from DFT to generate collective multi-particle moves for our lattice model.

There are, of course, other approaches that one can take to model the behaviour of liquid droplets on surfaces: Fully microscopic (particle resolved) approaches include (i) applying molecular dynamics computer simulations to solve Newton's equations of motion for all the individual atoms or molecules in the fluid -- see e.g.\ Ref.~\cite{tretyakov2013}, or (ii) Monte Carlo simulations -- see e.g.\ \cite{macdowell2002droplets}. As mentioned above, (iii) microscopic DFT can be applied, such as in Ref.~\cite{hughes2015liquid}, or DFT can also be applied to develop coarse-grained mesoscopic theories such as those described in Refs.~\cite{thiele2009, robbins2011modelling, chalmers2017dynamical}. Thin-film equation based models are also widely used \cite{de2013capillarity, oron1997long, kalliadasis2007, thiele2007structure, craster2009dynamics} and if thermal fluctuations are important, these can also be incorporated, creating stochastic thin-film equation models \cite{grun2006thin, duran2019instability}. Macroscopic scale approaches include just directly solving the Navier-Stokes equations or via methods such as lattice-Boltzmann \cite{kruger2017lattice}. However, when implementing such methods it is hard to connect to microscopic system properties (i.e.\ the underlying Hamiltonian), as we do here.

This paper is structured as follows: In Sec.~\ref{sec:2}, we introduce the lattice Hamiltonian that defines the energetics the model. In Sec.~\ref{sec:3}, the Monte Carlo single-particle dynamics algorithm and the thin-film evolution equation used to describe the effects of collective advective droplet dynamics over solid substrates are described. In Sec.~\ref{sec:DFT}, the DFT based statistical mechanical theory used for calculating the binding potential and other thermodynamic quantities is presented. We present results for the bulk fluid phase diagram, the liquid-vapour interfacial tension and the binding potential, showing how all of these depend on the temperature, on the fluid interaction parameters and the parameter governing the strength of the attraction of the fluid to the wall. In Sec.~\ref{sec:5}, we present equilibrium droplet properties, including calculating droplet density profiles using KMC and comparing the contact angle of droplets with the predictions from DFT. In Sec.~\ref{sec:6}, we present results from simulating nonequilibrium droplets on surfaces. This includes droplets joining via evaporation and condensation process, droplets sliding together and joining and also droplets sliding under lateral driving, e.g.\ due to gravity. Finally, in Sec.~\ref{sec:7}, we draw our conclusions.

\section{Lattice model and Hamiltonian}\label{sec:2}

\begin{figure}
\includegraphics[width=20pc]{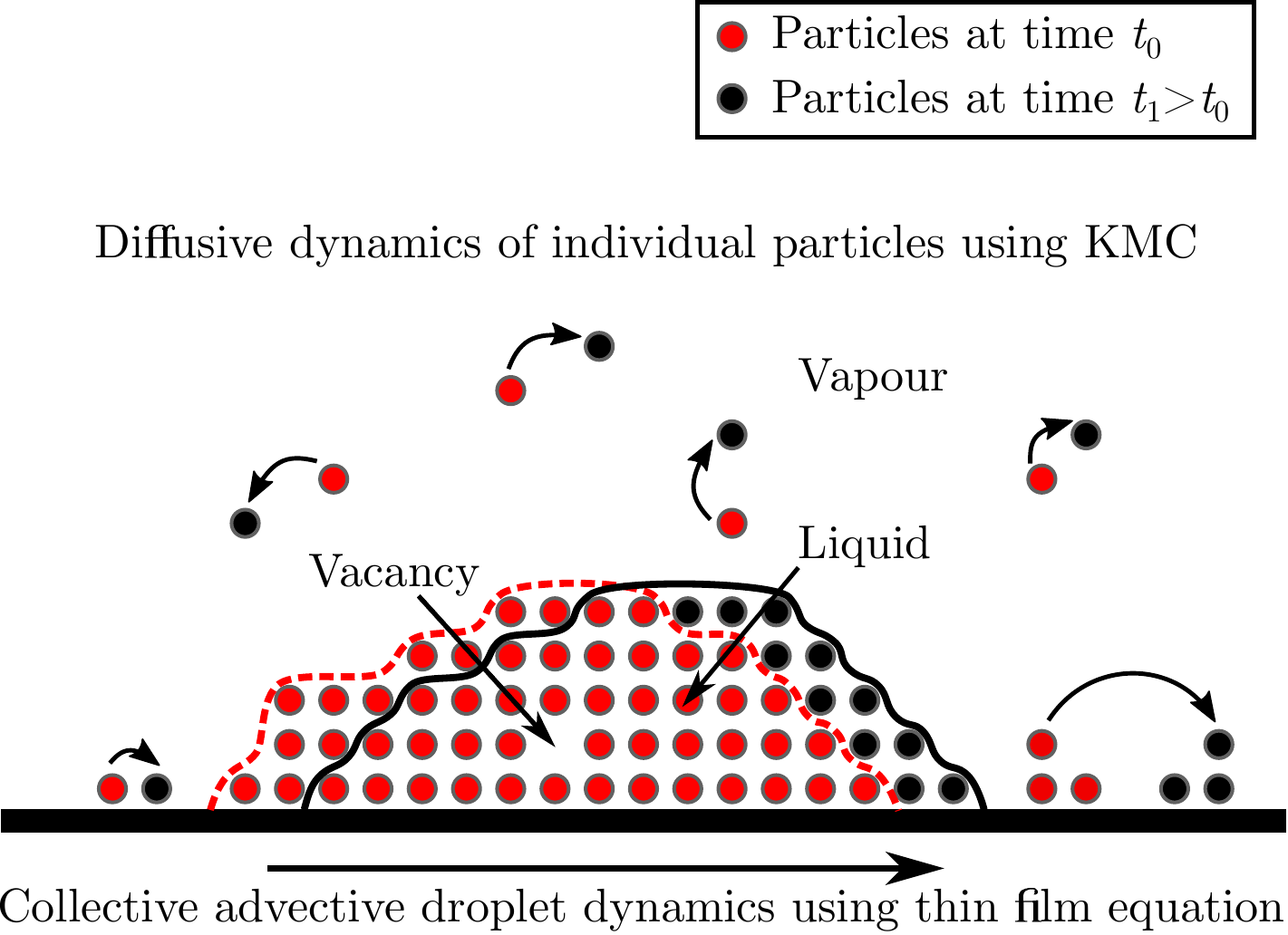}
\caption{A schematic of the system and how we model it: the liquid is discretised in space by setting it on a lattice. The dynamics is governed by a combination of single particle diffusive dynamics implemented by a KMC algorithm and collective advective droplet dynamics determined  by coupling to a thin-film equation.}
\label{fig:1}
\end{figure} 

In Fig.\ \ref{fig:1} is illustrated the system of interest and how it is modelled here. It consists of a planar solid surface upon which are a collection of interacting particles, with the majority gathered together forming a liquid droplet. We consider a Cartesian coordinate system $(x,y,z)$ with the $x$- and $y$-axis being parallel to the solid surface and the $z$-axis being perpendicular to the solid surface. The system is discretised onto a regular grid. To simplify, here we treat the system as being two-dimensional (2D), so that there are no variations in the $y$ direction, but the model can easily be generalised to three dimensions (3D). We define $\sigma$ to be the lattice spacing, which we set to $\sigma=1$, throughout. We define $l_{\bold{i}}$ to be the occupation number of lattice site $\bold{i}$,  where $\bold{i}=(i,j)$ is the 2D discrete position vector. We assume that $1\leq i\leq N_x$ and $0\leq j\leq N_y$, with $j=0$ corresponding to the solid surface. When $l_{\bold{i}}=1$ then site $\bold{i}$ is occupied by a particle and when it is unoccupied, then $l_\bold{i}=0$. We typically initiate the liquid in one or two semi-circular droplet configurations on the surface.

Note that one need not necessarily assume that each lattice site is of the same size as the individual molecules. One can also consider the model to be a coarse grained model, with $\sigma$ being much larger than the diameter of the molecules. In this case, when we refer to a lattice site as being `occupied by a particle' we mean that it contains mostly liquid and when a lattice site is referred to as `unoccupied' or `not containing a particle', then we mean that it mostly contains vapour. 

The total energy of the system $E$ is modelled by the Hamiltonian
\begin{equation}
\label{eq:Hamiltonian}
E= - \varepsilon\sum_{ \bold{i},\bold{j}}c _{\bold{ i},\bold{j}} l_{\bold{i}} l_{\bold{j}} +\sum_{\bold{i}}V_{\bold{i}}l_{\bold{i}},
\end {equation}
where the first term corresponds to a sum over pairs of lattice sites, and the overall strength of the interaction between particles is defined by the parameter $\varepsilon$. The interaction energy between pairs of particles at lattice sites $\bold{i}$ and $\bold{j}$ depends on the distance between them and is given by $c_{\bold{ i},\bold{j}}$. Here, we use 
\begin{equation}\label{eq:c_ij}
c_{\bold{i},\bold{j}}  = 
  \begin{cases} 
   1 & \text{if }\bold{j}\in {NN \bold{i}}, \\
    \frac{1}{2}    & \text{if }\bold{j}\in {NNN \bold{i}}, \\
    0   & \text{otherwise,}
         \end{cases}
\end{equation}
where $NN\bold{i}$ and $NNN\bold{i}$ denote the nearest neighbours of $\bold{i}$ and the next nearest neighbours of $\bold{i}$, respectively. The choice of values in Eq.\ (\ref{eq:c_ij}) is important, since with these any equilibrium droplets that form on the surface tend to have the shape in the form of a segment of a circle \cite{chalmers2017modelling, archer2010dynamical, robbins2011modelling}. If, for example, one were to assume only nearest neighbour interactions, i.e.\ with $c_{\bold{i},\bold{j}}=0$, for $\bold{j} \in {NNN \bold{i}}$, then rectangular shaped droplets are liable to be formed, particularly at low temperatures. The choice of $c_{\bold{i},\bold{j}}$ in Eq.\ (\ref{eq:c_ij}) minimises the dependence of the vapour-liquid surface tension on the interface-orientation. The values of $c_{\bold{i},\bold{j}}$ in Eq.\ (\ref{eq:c_ij}) come from considering how to discretise the Laplacian whilst minimising errors due to the discretisation \cite{archer2010dynamical, robbins2011modelling}.

In the second term in Eq.\ (\ref{eq:Hamiltonian}), is a sum over all the lattice sites and is the contribution to the potential energy from the external potential $V_{\bold{i}}$ due to  the surface on which liquid is deposited. We model it as follows:
\begin{equation}
\label{eq:ext_pot}
V_{\bold{i}}= 
  \begin{cases} 
    \infty  & \text{if }\bold{i}= (i,0),\\
    -\varepsilon_w & \text{if }\bold{i}= (i,1),\\
    0   &  \text{if } \bold{i}= \text{otherwise},
        \end{cases}
\end{equation}
where $\varepsilon_w$ is the parameter which determines the interaction strength between the particles and the solid surface (the wall). With this potential, the boundary condition for the particles at the wall is straight-forward: we set $l_{\bold{i}}= 0$ for all lattice sites with ${j} < 1$  in Eq.\ (\ref{eq:ext_pot}). As regards the other boundaries, for some of the simulations we use periodic boundary conditions on the left and right  boundaries that are perpendicular to the wall and for other simulations we make these hard purely repulsive walls. For the top boundary, we either assume that the system is closed, i.e.\ so that this is also a hard purely repulsive wall, or that the system is open, so that it is an absorbing boundary.

We also consider cases where there is a constant force ${\cal G}$ on the particles parallel to the surface, for example due to gravity. In this case, a term $j{\cal G}$ is added to the expression for the external potential $V_{\bold{i}}$ in Eq.~\eqref{eq:ext_pot}.

\section{Dynamics of the liquid}\label{sec:3}

Having defined the Hamiltonian, we must now specify the dynamics of the system. We use a combination of two types of particle moves to evolve the system forward in time: (i) Single particle moves, generated via the Metropolis Monte Carlo algorithm and (ii) collective many-particle moves, generated by considering the dynamics from a thin-film equation. We describe the moves of type (i) first.

\subsection{Monte Carlo dynamics}\label{subsec:KMC}

The Metropolis Monte Carlo algorithm \cite{landau2014guide} is commonly used to evolve systems in time, since it generates configurations with the correct (Boltzmann) equilibrium distribution:
 \begin{equation}
P_n=\frac{1}{Z}  e^{-\beta E_{n}},
\label{eq:equilib_P}
\end{equation}
where the integer index $n$ labels the sequence of configurations $\{l_{\bold{i}}\}$ generated by the algorithm, each having energy $E_{n} = E\left(\{l_{\bold{i}}\}\right)$, given by the Hamiltonian (\ref{eq:Hamiltonian}). The normalisation factor $Z$ is the partition function and $\beta = 1/k_{B}T$, where $k_B$ is Boltzmann's constant and $T$ is the temperature. The master equation governing the time evolution of the non-equilibrium probability $P_{n}(t)$ is \cite{landau2014guide}:
\begin{equation}
\label{eq:master}
\frac{\partial P_n(t) } {\partial t}= - \sum_{m \neq n}[P_n(t)W_{n \to m}-P_m(t)W_{m \to n}],
\end{equation}
where $W_{n \to m}$ is the transition probability for the system to go from state $n$ to state $m$. Eq.\ (\ref{eq:master}) has the form of a continuity equation, so the total probability remains normalised $\left(\sum_{n} P_n(t) \equiv 1\right)$ for all time. At equilibrium, where $\partial P_n/\partial t=0$, Eq.\ (\ref{eq:master}) yields
\begin{equation}
P_n(t)W_{n \to m} = P_m(t)W_{m \to n}.
\end{equation}
Since the equilibrium probability distribution function for state $n$ is given by Eq.\ (\ref{eq:equilib_P}), and similarly for state $m$, the ratio of the probabilities also gives the transition rate ratio
\begin{equation}
\label{eq:rate_ratio}
\frac{P_n(t)}{P_m(t)}=\frac{W_{m \to n}}{W_{n \to m}}= e^{-\beta\Delta E},
\end{equation}
where $\Delta E= E_n-E_m$. Below is outlined the Metropolis Monte Carlo algorithm that we implement, which samples with the above ratio of probabilities:
 \begin{enumerate}
\item Pick a random particle.
\item Pick randomly a nearest neighbour lattice site to the selected particle.
\item Check whether the neighbouring site is empty. If it is, calculate the change in energy $\Delta E$ using Eq.\ (\ref{eq:Hamiltonian}) corresponding to moving the particle into the empty site.
\item If $\Delta E<0$, then move the particle to its new position.
\item If $\Delta E>0$, allow the move with probability $e^{-\beta\Delta E}$. 
\end{enumerate}
The assumption we make (the standard assumption in KMC algorithms) is that Eq.\ (\ref{eq:rate_ratio}) still holds out of equilibrium and so we can use the above algorithm to evolve the system forward in time. The KMC algorithm outlined above leads to the particles performing a diffusive dynamics that at equilibrium generates configurations with the (correct) probability distribution \eqref{eq:equilib_P}.

\subsection{Hydrodynamics via the thin-film equation}

While the above KMC algorithm correctly samples the equilibrium states with a diffusive dynamics, this is not in fact what non-equilibrium liquids on surfaces actually do. We therefore introduce additional particle  moves to better model the hydrodynamics of the liquid on the surface. The additional particle moves are collective moves, transferring several particles simultaneously. These are governed by considering a thin-film equation from hydrodynamic theory \cite{thiele2007structure}. The effects of collective advective droplet hydrodynamic motion is illustrated in Fig.\ \ref{fig:1}. The film height profile $h(x,t_0)$ for the non-equilibrium droplet at time $t_0$ is given by the red dashed line, while the black solid line is the new height profile $h(x,t_1)$ that the droplet assumes at the later time $t_1$. In 3D, the thin-film equation is:
\begin{equation}
\label{eq:thin}
\frac{\partial h}{\partial t}=-\nabla\cdot\left[\frac{h^3}{3\eta}\left(\nabla[\gamma_{lv}\nabla^2 h+\Pi(h)]+\mathcal{G}\right)\right],
\end{equation}
where $ h(x,y,t)$ is the film thickness profile, $\gamma_{lv}$ is the liquid-vapour interfacial tension, $\eta$ is the viscosity and  $ \nabla=({\partial}/{\partial x},{\partial}/{\partial y})$. The term $ \gamma_{lv}\nabla^2 h$ represents the Laplace or  curvature pressure. Here, where we assume the system is 2D, we simply replace $\nabla\to{\partial}/{\partial x}$. $\mathcal{G}$ represents an additional force that may be present (e.g. gravity) acting in the positive $x$ direction. The term $\Pi(h)$ is the disjoining pressure, which describes the effective molecular forces between the substrate and the liquid-vapour interface. The disjoining pressure can be written as the derivative of an excess free energy \cite{de2013capillarity}, referred to as the binding potential, $g(h)$, as follows:
\begin{equation}
\label{eq:disj_press}
\Pi(h)= - \frac{\partial g}{\partial h}.
\end{equation}
The disjoining pressure is in fact a constrained excess free energy.  The equilibrium film thickness is given by the minimum value of $g(h_{min})$, where $h_{min}$ is the value of $h$ that minimises $g(h)$. When the minimum is at a small and finite value, then the liquid is said to be partially wetting. In contrast, when the minimum is at $h\to\infty$, then the liquid is said to be wetting. Additionally, the equilibrium contact angle $\theta$ is determined by the value of $g(h_{min})$, as follows \cite{churaev1995contact, rauscher2008wetting, hughes2015liquid}:
\begin{equation}
\label{eq:g_min_theta}
\theta=\cos^{-1}\left(1+\frac{g(h_{min})}{\gamma_{lv}}\right).
\end{equation}
In the literature, there are many different approximations used for the disjoining pressure $\Pi(h)$, depending on the nature of the fluid and the interactions with the surface \cite{de2013capillarity, thinfilms, israelachvili2011intermolecular, yin2017films}. Here, we use the binding potential that we calculate from DFT using the approach of Ref.~\cite{hughes2015liquid}. The form of $g(h)$ depends on the state point, i.e.\ on the particular values of $\varepsilon$, $\varepsilon_w$ and the temperature $T$. The DFT we use is also described below in Sec.\ \ref{sec:DFT}. The DFT data is fitted to the form \cite{hughes2015liquid, hughes2017, yin2017films}:
\begin{equation}
g(h)= a_1\sin(2\pi a_2 h+ a_3)+a_4\exp(-h/b)
+a_5\exp(-2h/b)+a_6\exp(-3h/b),
\label{eq:binding_pot}
\end{equation}
where $a_1$, $a_2$, $a_3$, $a_4$, $a_5$, $a_6$, and $b$ are constants that depend on the state point. We also use the DFT to calculate the liquid-vapour interfacial tension $\gamma_{lv}$, as described below in Sec.\ \ref{sec:DFT}. These are then used together with Eqs.\ (\ref{eq:thin}) and (\ref{eq:disj_press}) to determine the liquid droplet dynamics.

The relation between the KMC and thin-film steps is as follows. After completing a prescribed number of KMC steps, $M_1$, we first determine the discretised liquid film height profile from the configuration of occupied lattice sites. This is done by computing for each $i$ the number of occupied lattice sites above the lattice site $(i,0)$ up to a point where we find a vacancy of a certain number (we normally choose two) of consecutive lattice sites. This determines the height profile $h(x_i,0)$, where $x_i=i$. If for a certain $x_i$ it turns out that $h(x_i,0)=0$, we set it equal to the precursor film thickness $h_{min}$. This profile is then used as an initial condition for the thin-film equation. We solve the thin-film equation numerically using finite-difference approximations to represent spatial derivatives, and evolve the solution in time using e.g. Euler's method. We choose a time step, $\Delta t$, and solve the thin-film equation for a certain number of time steps, $M_2$. Next, we map the computed thin-film profile back to the lattice by appropriately removing/adding particles where the film profile became thicker/thinner, as illustrated in Fig.~\ref{fig:1}. Note that this does not change the position of any particles in the vapour phase. To ensure the conservation of the number of the particles in the system, we then count the total number of the particles and add or remove a few particles, if needed, at random positions above the computed thin-film profile. The resulting lattice site configuration can then be used again for the KMC computations. The evolution of the system is obtained by successively repeating the KMC and thin-film steps as described above for a certain number of times, $n$.

\subsection{Diffusion coefficient and time scales}

An important quantity in the procedure described above is the parameter $\lambda$ proportional to the ratio of the number of KMC attempted particle moves per particle $M_1/N$ to the number of thin-film moves $M_2$, or more strictly the total time the thin film equation is evolved for at each step, $M_2\Delta t$. In dimensionless time units, this ratio becomes:
\begin{equation}\label{eq:lambda}
\lambda=\frac{M_1\tau}{M_2 N \Delta t}.
\end{equation}
Here, $N$ is the total number of the particles in the system and $\tau$ is proportional to the viscosity of the liquid, $\eta$:
 \begin{equation}
 \tau=3 \eta \beta \sigma^3. \label{eq:tau}
 \end{equation}
When $\lambda=0$, the system is evolved by using only the thin-film equation, whilst the limit $\lambda\rightarrow\infty$ corresponds to evolving the system using only KMC particle moves. {Since the single-particle KMC moves generate diffusive type motion, a large $M_1/N$ leads to a higher diffusion coefficient.} Therefore, $\lambda\propto D/\eta$, where $D$ is a single-particle diffusion coefficient. The bulk liquid diffusion coeficient for a 3D model very similar to that considered here, with dynamics governed by the same KMC algorithm, was determined in Ref.~\cite{chalmers2017modelling}. For temperatures in a similar range to those considered here, they obtained the result that $D\approx2.6\times10^{-4}\sigma^2$ KMC steps$^{-1}$, where $x$ KMC steps means $x$ attempted moves per lattice site. In the present 2D system, one should expect $D$ to be of a similar magnitude in the bulk liquid. As regards the diffusion coefficient for isolated particles on the surface, in this regime $D\sim\sigma^2\lambda\propto M_1$. {Recalling the Einstein relation $D=(6\pi\sigma\eta\beta)^{-1}$, this may be used if seeking to match results from the present model to particular experimental systems. Moreover, this allows to see that the matched value of $\lambda$ is largely determined by the viscosity and the temperature.}


 \section{Lattice DFT for the fluid}
\label{sec:DFT}

We use classical DFT \cite{evans1979nature, hansen2013theory} adapted to lattice models \cite{hughes2014introduction}, in order to calculate the binding potential $g(h)$ and also the liquid-vapour interfacial tension $\gamma_{lv}$. These quantities are then input into the thin-film equation to create the collective hydrodynamic moves, as described in the previous section. We also use the DFT to calculate the bulk fluid phase diagram.

DFT is a statistical mechanical theory for the fluid one-body average density profile, which on a lattice is the quantity:
\begin{equation}
\rho_{\bold{i}}=\langle l_{\bold{i}} \rangle,
\label{eq:rho_def}
\end{equation}
where $\langle\, \boldsymbol{\cdot}\, \rangle$ denotes an ensemble average. The approximation for the Helmholtz free energy that we use is
\begin{equation}
F(\{\rho_\bold{i}\})=k_{B}T \sum_{ \bold{i}} [\rho_{\bold{i}}\ln\rho_{\bold{i}} + (1 - \rho_{\bold{i}})\ln(1 - \rho_{\bold{i}})]
- \varepsilon \sum_{ \bold{i},\bold{j}} c _{\bold{ i},\bold{j}} \rho_{\bold{i}} \rho_{\bold{j}}  + \sum_{\bold{i}}V_{\bold{i}} \rho_{\bold{i}}. 
\label{eq:free_energy}
\end{equation}
A derivation of this is given in Ref.~\cite{hughes2014introduction}. There is an obvious (mean-field) connection between the Hamiltonian \eqref{eq:Hamiltonian} and the terms on the second line in Eq.\ \eqref{eq:free_energy}, and the first term is essentially entropic in origin \cite{hughes2014introduction}. The equilibrium fluid density profile $\{\rho_\bold{i}\}$ is obtained by minimising the grand potential
\begin{equation}
\Omega(\{\rho_\bold{i}\})= F(\{\rho_\bold{i}\}) -\mu \sum_\bold{i}\rho_\bold{i}, 
\label{eq:grand_pot}
\end{equation}
where $\mu$ is the chemical potential \cite{evans1979nature, hansen2013theory}. 

\subsection{Bulk fluid phase diagram from DFT}

\begin{figure}
\includegraphics[width=20pc]{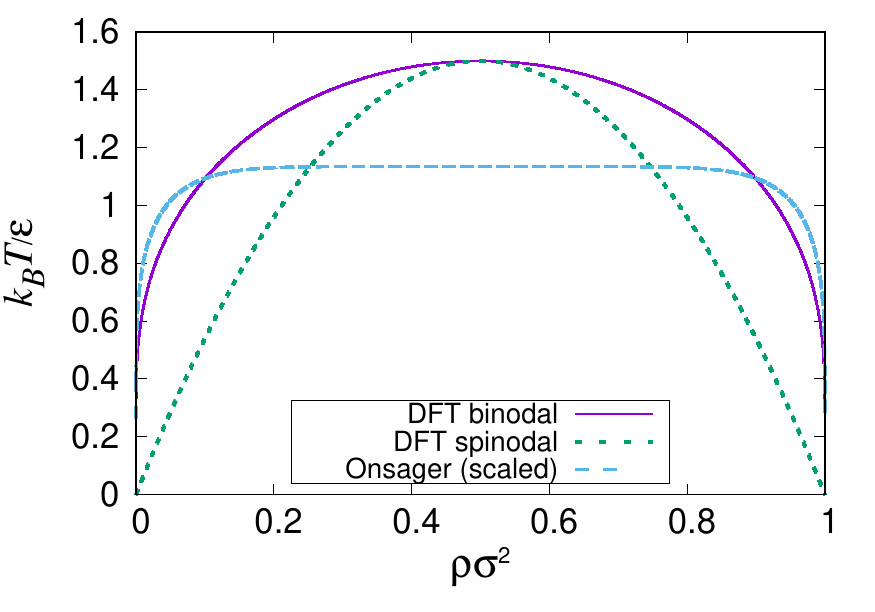}
\caption{The bulk fluid phase diagram in the temperature versus density plane obtained from DFT together with the Onsager result for the 2D Ising model scaled to match the observed critical temperature. The binodal gives the densities of the coexisting vapour and liquid phases as the temperature is varied. The spinodal is also displayed.}
 \label{fig:phase_diag}
\end{figure} 

Determining the phase diagram of the bulk fluid, away from the influence of any interfaces, is a prerequisite to embarking on any study of the behaviour of the fluid at walls. Phase coexistence between a high density liquid phase and a low density vapour phase is exhibited when the temperature $T$ is less than the critical temperature $T_c$. Thermodynamic coexistence of two phases occurs when the temperature, chemical potential and pressure are all equal in the two phases. The locus in the phase diagram of these coexisting states is referred to as the binodal and for our model is displayed in Fig.\ $\ref {fig:phase_diag}$, plotted in the density versus temperature plane. We display the result from the DFT \eqref{eq:free_energy} (see the following paragraph for the details of how this is calculated) together with an estimate for the true location of the binodal. The latter is made by determining $T_c$ using our Monte Carlo simulations, based on Eq.~(\ref{eq:Hamiltonian}), and then by scaling the exact Onsager 2D Ising result for the binodal \cite{baxter2012onsager}, so that the critical temperature matches that from our simulations, $T_c\approx1.1\varepsilon/k_B$. Recall that the Onsager result is for the standard 2D Ising which just has nearest neighbour interactions, i.e.\ setting $c_{\bold{i},\bold{j}} = 0$ for $\bold{j}\in{NNN\bold{i}}$. We see that in the region near $T_c$, the mean-field DFT fails, but for temperatures ${k_B}T/ \varepsilon < 1$ the agreement between the DFT and the Monte Carlo results is fairly good, as we also see  below where we present results for droplets on surfaces and for the wetting behaviour.

To calculate the binodal predicted by the free energy \eqref{eq:free_energy}, a symmetry in the lattice-gas model proves to be rather useful. Note that this symmetry is absent in continuum models of fluids. If we replace $l_\bold{i} = 1 - o_\bold{i}$ in the Hamiltonian (\ref{eq:Hamiltonian}), then we see that this leaves the form of it unchanged. We can think of $o_\bold{i}$ as being a `hole' occupation number. As a consequence of this particle-hole symmetry, we therefore have the following relation between the density of the liquid $\rho_l$ and the vapour $\rho_v$ at coexistence:
\begin{equation}
\label{eq:symmetry_relation}
\rho_l=1-\rho_v.
\end{equation}
From Eq.\ (\ref{eq:free_energy}), we obtain the Helmholtz free energy per unit area in the absence of any external field:
\begin{equation}
\label{eq:free_energy_uniform}
f = \frac{F}{A} = k_{B}T [\rho\ln\rho+ (1 - \rho)\ln(1 - \rho)] -3 \varepsilon\rho^2,
\end{equation}
where $A$ is the area of the system. Recall that we are considering a 2D system; one should, of course, consider the free energy per unit volume $F/V$ in 3D, where $V$ is the volume. Note also that we have used the result $\varepsilon \sum_{ \bold{i},\bold{j}} c _{\bold{ i},\bold{j}}=6\varepsilon/2$, to obtain Eq.\ \eqref{eq:free_energy_uniform}. This is the sum over the interactions of a particle with all of its neighbours [see Eq.~\eqref{eq:c_ij}]. The factor $1/2$ is present to avoid double counting the interactions between each of the pairs. The pressure in the system is therefore \cite{hansen2013theory}
\begin{equation}
\begin{aligned}
p&=-\left(\frac{\partial F}{\partial V}\right)_{T,N}=\rho\dfrac{\partial f}{\partial \rho} -f\\
&= - k_{B}T \ln(1 - \rho) -3\varepsilon\rho^2.
\end{aligned}
\end{equation}
Invoking Eq.\ (\ref{eq:symmetry_relation}) and solving $p(\rho)=p(1-\rho)$  leads to the binodal curve
\begin{equation}
\label{eq:binodal}
\frac{k_BT}{\varepsilon}=\dfrac{3(2\rho-1)}{\ln (\rho/(1-\rho))}.
\end{equation}
This is the curve displayed in Fig.\ $\ref {fig:phase_diag}$. The maximum on the binodal corresponds to the critical point temperature, above which there is no vapour-liquid phase separation. Substituting the critical density $\rho=1/2$ into Eq.~\eqref{eq:binodal} gives the critical temperature predicted by the DFT to be $T_c=1.5\varepsilon/k_B$, while from our KMC simulations we find that in reality it is $T_c \approx1.1\varepsilon/k_B$. We can also calculate the chemical potential 
\begin{equation}
\begin{aligned}
\label{eq:mu}
\mu&=-\bigg(\frac{\partial F}{\partial N}\bigg)_{T,V}=\dfrac{\partial f}{\partial \rho}\\
&=k_{B}T \ln\bigg(\dfrac {\rho}{1-\rho}\bigg) - 6\varepsilon\rho.
\end{aligned}
\end{equation}  
The value of the chemical potential at bulk phase coexistence (i.e.\ on the binodal) is found by substituting Eq.\ (\ref{eq:binodal}) into Eq.\ (\ref{eq:mu}), which gives 
\begin{equation}
\mu_{coex}=-3\varepsilon.
\label{eq:mu_coex}
\end{equation}
In Fig.\ $\ref{fig:phase_diag}$ we also display the spinodal predicted by the DFT. Spontaneous phase separation occurs within this region of the phase diagram, since the uniform fluid is unstable for state points within the spinodal curve. The condition to determine the spinodal is
\begin{equation}
\label{eq:spinodal_condition}
\dfrac{\partial^2f}{\partial \rho^2}=0.
\end{equation}
Using this together with Eq.\ \eqref{eq:free_energy_uniform} we obtain the following result for the spinodal
\begin{equation}
\label{eq:spinodal}
\frac{k_BT}{\varepsilon}=6\rho(1-\rho),
\end{equation}
which is the curve plotted in Fig.\ \ref{fig:phase_diag}. {Note that the particle-hole symmetry discussed above is manifested in all the curves in Fig.\ \ref{fig:phase_diag}. Of course, real continuum liquids do not have this symmetry. This symmetry can be broken by introducing a 3-body term $\propto\sum_{ \bold{i},\bold{j},\bold{k}}\tilde{c} _{\bold{ i},\bold{j},\bold{k}}l_{\bold{i}} l_{\bold{j}}l_{\bold{k}}$ in the Hamiltonian \eqref{eq:Hamiltonian} -- see Ref.~\cite{MB85} for more details. We do not pursue this possible extension to the model here.}

\subsection{Properties of the vapour-liquid interface}

\begin{figure}[t]
\center
\includegraphics[width=20pc]{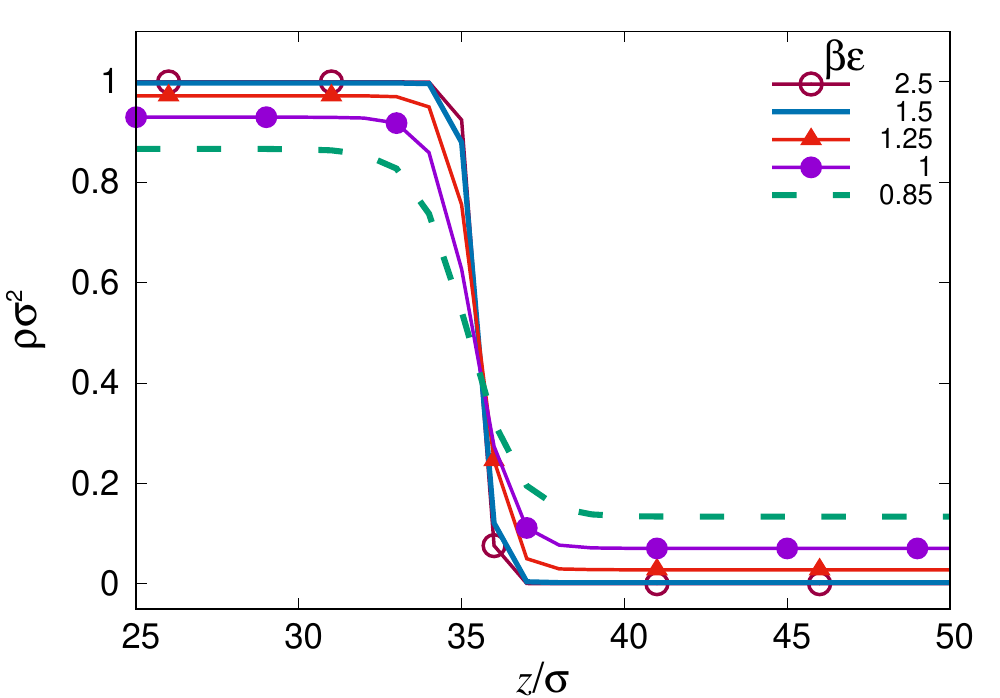}
\caption{A series of density profiles for the vapour-liquid interface obtained from the DFT (\ref{eq:free_energy}) for various values of $\beta{\varepsilon}$, as indicated in the key. These are used to calculate the vapour-liquid interfacial tension.}
\label{fig:free_interface}
\end{figure}

Having determined the bulk fluid phase diagram, one can then determine the properties of the interface between the coexisting liquid and vapour phases. In Fig.\ \ref{fig:free_interface} we display a sequence of density profiles for the vapour-liquid interface, calculated over a range of different temperatures. The density profiles are calculated by minimising the grand potential (\ref{eq:grand_pot}) with the Helmholtz free energy $F$ given by Eq.\ (\ref{eq:free_energy}) and the chemical potential $\mu$ set to be the value at coexistence, given in Eq.\ (\ref{eq:mu_coex}). To calculate these we set the external potential $V_\bold{i}=0$ and assume that the profiles only vary in one direction (along the $z$-axis). We also assume the system has periodic boundary conditions and initiate the system with half of it containing the liquid and the other half containing the vapour. The final equilibrium therefore contains two vapour-liquid interfaces, but only one of these is displayed in Fig.\ \ref{fig:free_interface}.

\begin{figure}[t]
\center
\includegraphics[width=20pc]{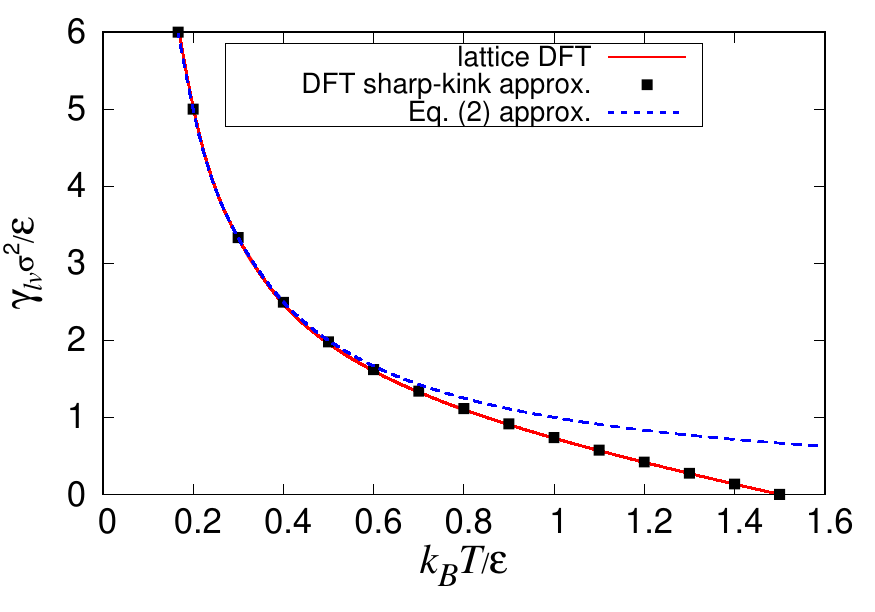}
\caption{The liquid-vapour interfacial tension as a function of temperature, calculated using our DFT (solid line), calculated using the sharp-kink approximation (symbols), and also calculated using Eq. (\ref{eq:Hamiltonian}) (dashed line), assuming a sharp interface and that the liquid has $l_\bold{i}=1$ and the vapour has $l_\bold{i}=0$.}
\label{fig:surface_tension}
\end{figure} 

Having calculated these profiles, we can then calculate the liquid-vapour interfacial tension $\gamma_{lv}$. Recall that a planar interface between the vapour and the liquid in a 3D system adds to the free energy of the system an excess (over bulk) contribution equal to the interfacial tension multiplied by the area of the interface \cite{evans1979nature, hansen2013theory}. Similarly, for the present 2D system the interfacial contribution is equal to $2L\gamma_{lv}$, where $2L$ is the length of the interface. The factor 2 arises because there are two interfaces in our system with periodic boundary conditions. Having calculate the grand potential $\Omega$ for the system containing the interface, the interfacial tension is then 
\begin{equation}
\gamma_{lv}=\dfrac{\Omega-\Omega_{0}}{2L},
\end{equation}
where
\begin{equation}
\Omega_{0}=-pA,
\label{eq:Omega_0}
\end{equation}
is the grand potential for a 2D box having the same area $A$, but only containing one of the two phases, and where $p$ is pressure. For example, when $\beta\varepsilon = 2.5$, this approach gives the vapour-liquid interfacial tension to be $\beta\gamma_{lv}= 2.48$.

In Fig.\ \ref{fig:surface_tension} we display the surface tension of the planar liquid-vapour interface plotted as a function of temperature. We see that as the temperature $T$ approaches from below the critical temperature $T_c=1.5\varepsilon/k_B$, then $\gamma_{lv}\rightarrow 0$, due to the fact that the difference in the density of the two coexisting phases goes to zero as $T\to T_c$. In addition to the full DFT results for $\gamma_{lv}$, in Fig.\ \ref{fig:surface_tension} we also display the results from a so-called `sharp-kink' approximation (c.f.\ \cite{dietrich88, stewart2005wetting}), where we assume that the interface consists of a sharp step from the liquid with density $\rho_l$, to the vapour with density $\rho_v$, both being the values at phase coexistence; see Eqs.~\eqref{eq:symmetry_relation} and \eqref{eq:binodal}. The excess free energy for this density profile is then evaluated via Eq.~\eqref{eq:free_energy}. We also show in Fig.~\ref{fig:surface_tension} the results from an even cruder sharp-kink type approximation that assumes that in the liquid $l_\bold{i}=1$  and then use the Hamiltonian to evaluate the energy change when such a system is cleaved in two to create an interface with a vacuum, i.e.\ we effectively assume for this calculation that in the vapour $l_\bold{i}=0$, with a step-like planar interface between the two. We see from Fig.\ \ref{fig:surface_tension} that there is good agreement between the value for the surface tension calculated from DFT and from the first sharp-kink approximation for all temperatures and also with the second sharp-kink approximation for temperatures $k_{B}T/\varepsilon < 0.8$, indicating that in this regime the major contribution to the interfacial tension is energetic, rather than entropic.

The calculations of the solid-vapour and solid-liquid interfacial tensions, $\gamma_{sv}$ and $\gamma_{sl}$, respectively, are done in a similar manner to the calculation of $\gamma_{lv}$ described above, but of course the wall potential is retained and the system is initialised with just either the vapour or the liquid density values, respectively. For additional details on how these calculations are done, see e.g.\ the discussion in Ref.~\cite{hughes2014introduction}. 

\subsection{Calculation of the binding potential using DFT}
\label{sec:DFT_binding_pot}

\begin{figure}[htp]
\includegraphics[width=20pc]{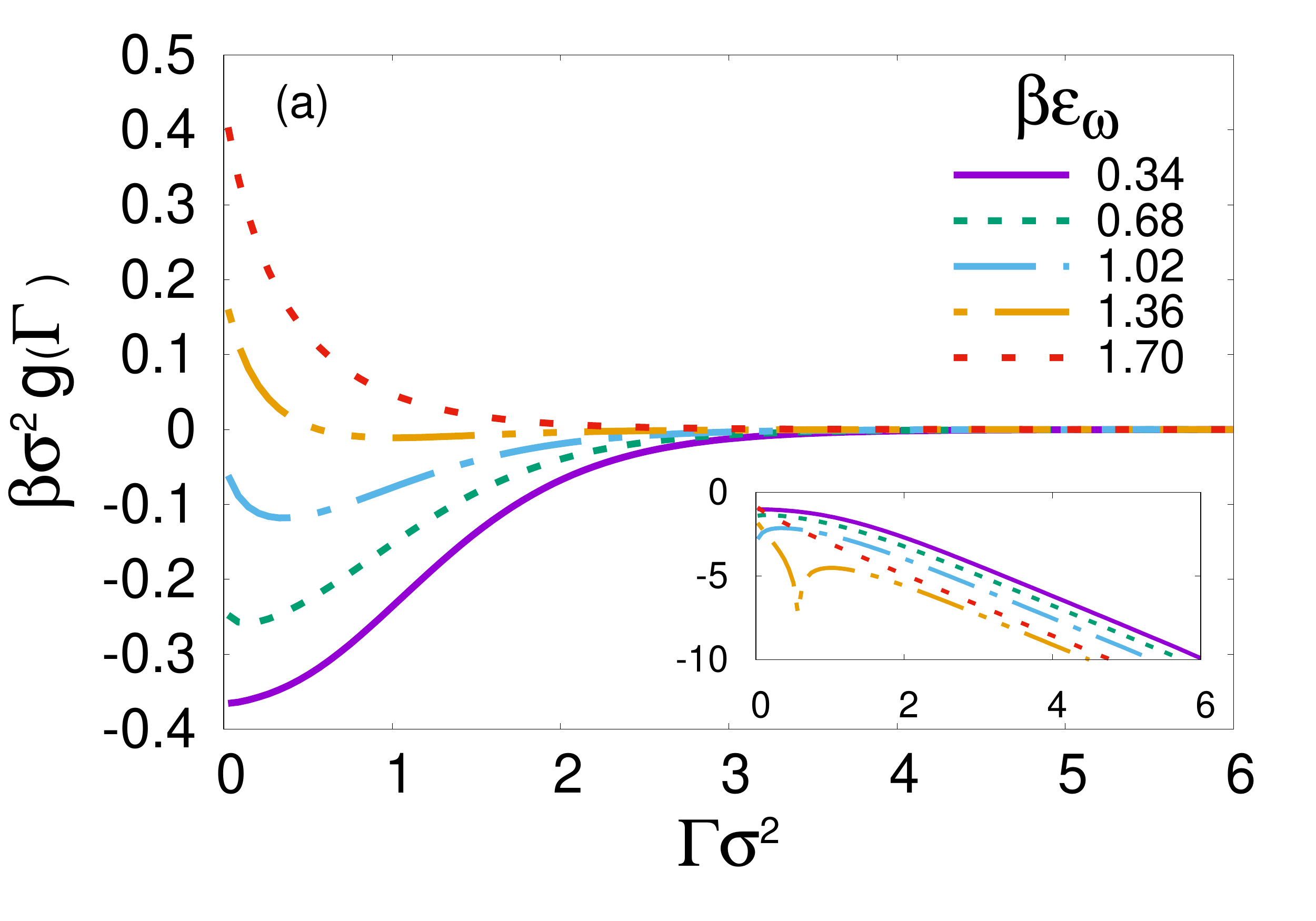}
\includegraphics[width=20pc]{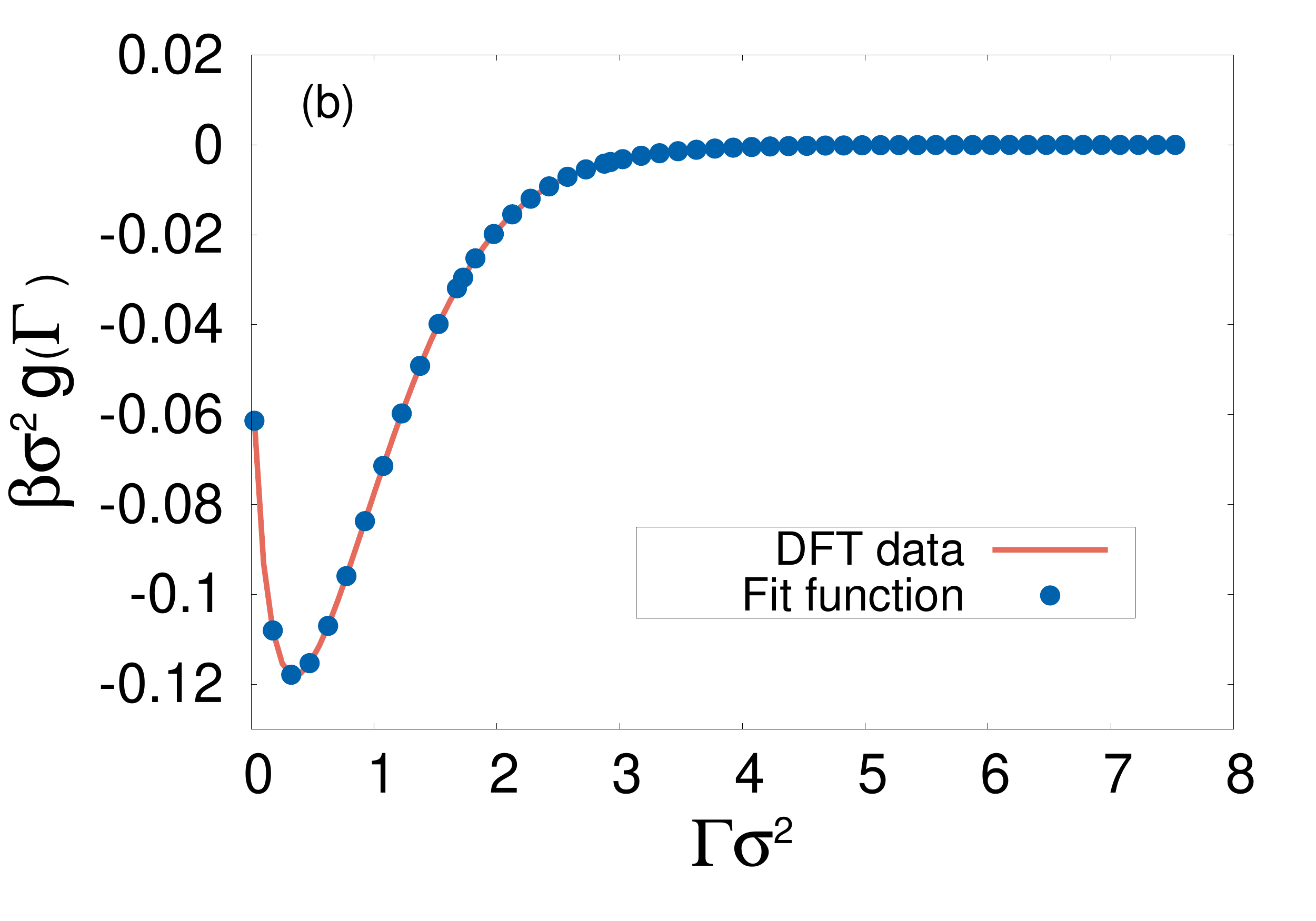}

\caption{The binding potential calculated from DFT for $\beta\varepsilon= 0.85$. In (a) are results for varying $\beta\varepsilon_w$, as given in the key. The inset shows the exponential asymptotic decay form for $\Gamma\to\infty$ in the tails of the binding potentials. In (b) we show a comparison of the DFT results with the fit function in Eq.\ (\ref{eq:binding_pot}) for $\beta\varepsilon_w = 1.02$.}
\label{fig:Binding0_85}
 \end{figure}
 
\begin{figure}[htp]
\includegraphics[width=20pc]{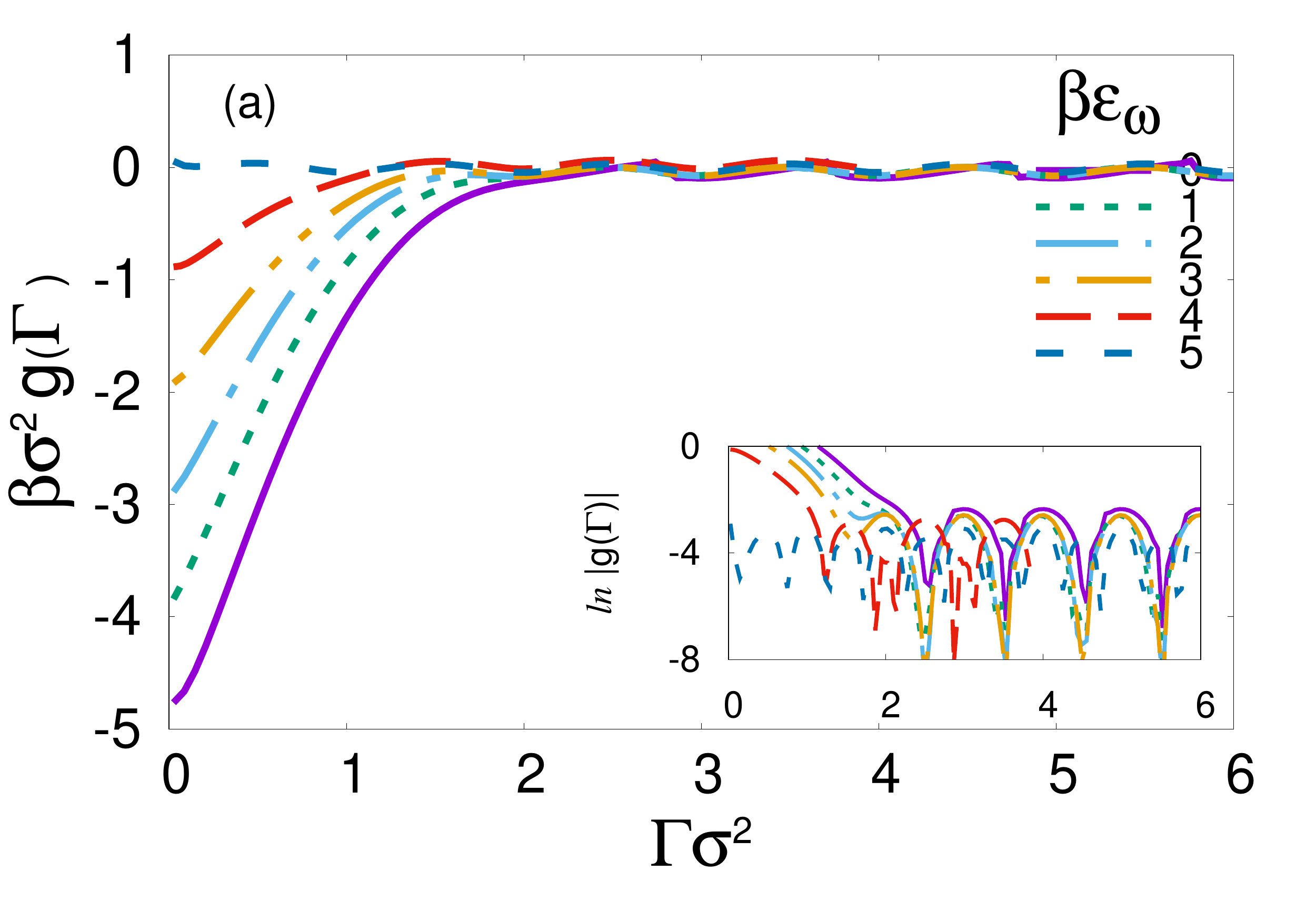}
\includegraphics[width=20pc]{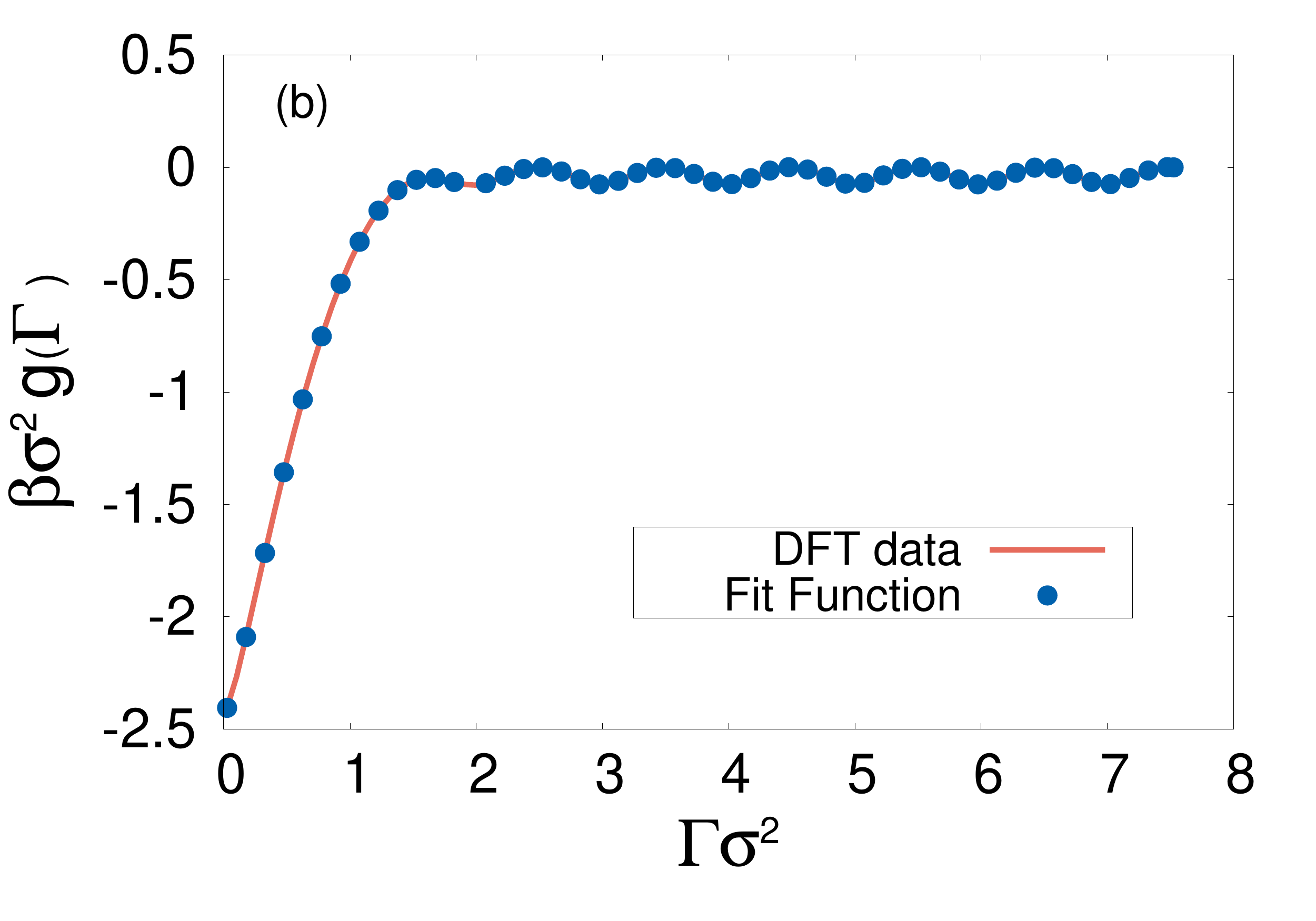}

\caption{The binding potential calculated from DFT for $\beta\varepsilon= 2.5$. In (a) are results for varying $\beta\varepsilon_w$, as given in the key. The inset shows the oscillations that occur in the tails of the potentials that arise as a result of the system being a lattice model. In (b) we show a comparison with the fit function in Eq.\ (\ref{eq:binding_pot}) for $\beta\varepsilon_w = 2.5$.}
\label{fig:Binding2_5}
\end{figure}

The excess grand potential per unit length (per unit area, in 3D) for a film of liquid adsorbed between a wall and a bulk vapour can be written as \cite{dietrich88, hughes2014introduction, hughes2015liquid, hughes2017}
\begin{equation}
\omega_{ex}(h)\equiv\frac{\Omega-\Omega_0}{L}=\gamma_{sl}+\gamma_{lv}+g(h)+\Gamma\delta\mu,
\label{eq:little_omega}
\end{equation}
where $L$ is the length of the interface with the wall, $\Omega_0$ is given by Eq.\ \eqref{eq:Omega_0}, $\delta\mu=(\mu_{coex}-\mu)$, $g(h)$ is the binding potential and $\Gamma$ is the adsorption, given by
\begin{equation}
\label{eq:Gamma_def}
\Gamma=\frac{1}{L}\sum_\bold{i}(\rho_\bold{i}-\rho_v),
\end{equation}
which is the excess (over bulk) number of particles adsorbed on the interface. Note that strictly we should consider $g$ to be a function of the adsorption $\Gamma$, not of $h$, since the latter is not a well defined quantity \cite{hughes2015liquid}. However, when $h$ is large Eq.~\eqref{eq:Gamma_def} gives
\begin{equation}
\Gamma \approx  h (\rho_l-\rho_v).
\label{eq:adsorption}
\end{equation}
Therefore, throughout this paper we discuss $h$ and $\Gamma$ almost interchangeably, and treat Eq.~\eqref{eq:adsorption} as an equality which defines the value of $h$.

When the system is at vapour-liquid phase coexistence we have $\delta\mu=0$ and so the last term in Eq.\ \eqref{eq:little_omega} is zero. We also see that when $h\to\infty$ we must have $g(h)\to0$, since when $\delta\mu=0$ the only contributions to $\omega_{ex}$ must be those from the wall-liquid and the liquid-vapour interfaces, i.e.\ $\gamma_{sl}$ and $\gamma_{lv}$, respectively. These considerations also enables us to see what $g(h)$ really is: it is the additional contribution to $\omega_{ex}$ from the interaction between these two interfaces when they become close to one another. For the case when the liquid is partially wetting, i.e.\ when $g(h)$ has a minimum at a finite value of $h$, then $g(h)$ `binds' these two interfaces together, hence the name given to $g(h)$. This interaction between the interfaces stems from the molecular interactions between fluid particles and with the wall and can be thought of as arising in a similar manner to the surface tension.

The equilibrium film thickness is that minimises Eq.~\eqref{eq:little_omega}. When $\delta\mu=0$, this is obtained by solving $\Pi(h)=-\frac{\partial g}{\partial h}=0$. For all other values of $h$, the system is out of equilibrium and therefore to determine $g(h)$ for other values of $h$, one must either move away from coexistence or apply a constraint, the constraint being that there is a specified thickness $h$ of liquid at the wall.

The approach used to calculate $g(h)$ here is that developed in Refs.~\cite{archer2011nucleation, hughes2015liquid}, which consists of a constrained free energy minimisation. This method uses an iterative algorithm to minimise the grand potential (\ref{eq:grand_pot}) subject to the constraint that the adsorption $\Gamma$ is a specified value, which via Eq.~\eqref{eq:adsorption} is equivalent to minimising subject to the constrain that the film thickness $h$ is a specified value. It turns out that this constraint is equivalent to applying a fictitious additional external potential that stabilises the liquid film of the specified thickness \cite{archer2011nucleation, hughes2015liquid}. The calculation is repeated for a sequence of different values of $\Gamma$, enabling us to calculate $g(\Gamma)$ over the whole range of values of $\Gamma$. This data set is subsequently fitted to the form given in Eq.\ \eqref{eq:binding_pot} and then input into the thin-film equation \eqref{eq:thin}.

In Figs.\ \ref{fig:Binding0_85} and \ref{fig:Binding2_5} we display results for the binding potential determined via this method for a range of different state points. In Fig.\ \ref{fig:Binding0_85} we display results for $\beta\varepsilon = 0.85$, which corresponds to a temperature that is not that far below the critical temperature $T_c$. In Fig.\ \ref{fig:Binding0_85}(a) are results for a sequence of different values of the wall attraction strength $\varepsilon_w$. We see that for small values of $\beta\varepsilon_w$, which corresponds to a weakly attracting solvophobic wall, the global minimum of $g(\Gamma)$ is at a small value of $\Gamma$, i.e.\ the liquid does not wet the wall. As $\beta\varepsilon_w$ is increased, there is a first order wetting transition when $\beta\varepsilon_w \approx 1.36$ and for $\beta\varepsilon_w > 1.36$, the global minimum in $g(\Gamma)$ is at $\Gamma \rightarrow \infty$; i.e.\ for these values of $\beta\varepsilon_w$ the liquid wets the wall. The inset of Fig.\ \ref{fig:Binding0_85}(a) shows the same results with the vertical axis plotted on a logarithmic scale to show more clearly the monotonic decay behaviour as $h\to\infty$. In Fig.\ \ref{fig:Binding0_85}(b) we show the result for $\beta\varepsilon_w=1.02$ to demonstrate the excellent agreement between the numerical results for $g(\Gamma)$ and the fit function in Eq.\ \eqref{eq:binding_pot}. This shows clearly that this simple form is highly appropriate for fitting the binding potential over the whole range of values of $\Gamma$ \cite{hughes2015liquid, hughes2017, tewes2017comparing}.

\begin{figure}
\floatbox[{\capbeside\thisfloatsetup{capbesideposition={left,top},capbesidewidth=15pc}}]{figure}[\FBwidth]
{\caption{\label{fig:densityprofile2_5} A series of 2D density profiles corresponding to a liquid droplet on a surface, calculated using KMC for $\beta{\varepsilon} = 2.5$ and for varying values of the wall attraction strength $\beta{{\varepsilon}_w}=0.5$, $1.5$, $2.5$, $3.5$, $4.5$ and $5$. These are obtained by averaging over a series of configurations generated by $4\times10^9$, $10\times10^9$, $2\times10^8$, $6\times10^9$, $9\times10^9$ and $20\times10^9$ attempted moves, respectively.}}
{\includegraphics[width=20pc]{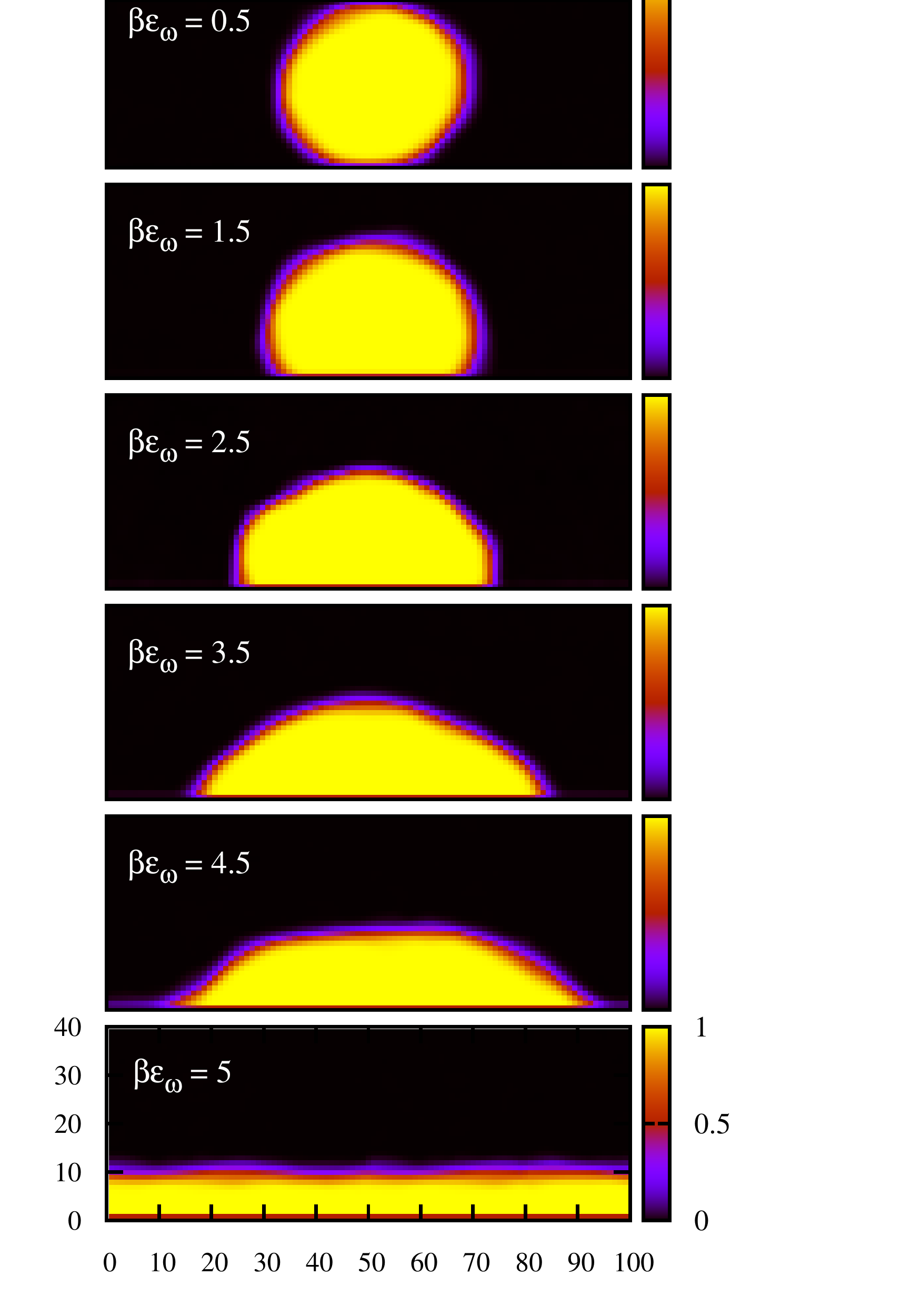}}
\end{figure}

In Fig.\ \ref{fig:Binding2_5} we display results for the binding potential for $\beta\varepsilon = 2.5$, which corresponds to a much lower temperature than the results in Fig.\ \ref{fig:Binding0_85}. The results in Fig.\ \ref{fig:Binding2_5}(a) are for a range of different values of $\varepsilon_w$ and we see that for $\beta\varepsilon_w\approx5$ the minimum in $g(\Gamma)$ close to the wall dissapears, so that for $\beta\varepsilon_w > 5 $ the global minimum in $g(\Gamma)$ is at a larger value of $\Gamma$. However, we note that in this case, due to the presence of oscillations in $g(\Gamma)$ that do not decay in amplitude, there is an infinite sequence of minima in $g(\Gamma)$ as $\Gamma\to\infty$. Oscillations that decay in amplitude can occur in the binding potential as a result of the nature of the molecular interactions \cite{hughes2017, yin2017films}. However, in the present system the oscillations are due the low temperature and the discrete (lattice) nature of the model \cite{hughes2015liquid}. Each subsequent minimum in $g(h)$ corresponds to the vapour-liquid interface moving by a distance of one lattice site further from the wall. The inset of Fig.\ \ref{fig:Binding2_5}(a) shows the same results with the vertical axis plotted on a logarithmic scale to show more clearly the oscillatory decay behaviour. We see in Fig.~\ref{fig:Binding2_5}(b) that, as for the previous case, the fit function \eqref{eq:binding_pot} gives an excellent representation of the numerical DFT results, in this case for $\beta\varepsilon_w=2.5$.


\section{Contact angle and static droplet profiles from KMC}\label{sec:5}


As described in the previous section, it is straightforward to calculate the three interfacial tensions $\gamma_{lv}$, $\gamma_{sl}$ and $\gamma_{sv}$ using our lattice DFT \eqref{eq:free_energy}. These can then be inserted into Young's equation \eqref{eq:Young} to determine the equilibrium contact angle $\theta$. This is also the contact angle that equilibrium droplet solutions of the thin-film equation have, since $\theta$ is determined by $g(h)$, as Eq.\ \eqref{eq:g_min_theta} shows. It is important that the value of $\theta$ determined by $g(h)$ in the thin-film equation is close to the value of the contact angle that the KMC simulations drive the system towards. If there is too big a difference between the contact angle from DFT and from the KMC simulations, then when we have both the KMC and thin-film dynamics in our model, these will work against each other, which would be unsatisfactory. Thus, the contact angle from the DFT and from the KMC must be close in value. This is also a check of the accuracy of the DFT. Of course, at higher temperatures near $T_c$ the DFT is not accurate; that is clear from the phase diagram in Fig.\ \ref{fig:phase_diag}. However, at lower temperatures we do find that the DFT becomes rather accurate, as we now show.

In Fig.\ \ref{fig:densityprofile2_5} we display a sequence of density profiles, each obtained by averaging over a time series from our KMC simulations, for various different values of the wall attraction parameter $\beta{{\varepsilon}_w}$ and for the temperature $k_BT/\varepsilon=1/2.5$. For these simulations we do not include any of the thin-film collective dynamics, since at this stage we are solely interested in equilibrium properties. In each case there are $N = 897$ particles in the system, which has a total domain size of $100 \times100$ lattice sites and with periodic boundary conditions between the left and right sides. We initiate the system by setting $l_\bold{i}=1$ for all the lattice sites within a semi-circular region in the middle upon the substrate and we set $l_\bold{i}=0$ outside the semi-circle. We see that for the largest value of the wall attraction $\beta{\varepsilon}_w=5$, the liquid spreads out over the substrate; i.e.\ it wets the substrate. For smaller values of $\beta{{\varepsilon}_w}$, the liquid forms a droplet on the substrate with an increasingly large contact angle as $\beta{{\varepsilon}_w}$ is decreased; i.e.\ for these cases the liquid is partially wetting. This behaviour and the observed wetting transition at $\beta{{\varepsilon}_w}\approx5$ is in very good agreement with what we expect based on Fig.~\ref{fig:Binding2_5}. Note too that the droplet profiles in Fig.\ \ref{fig:densityprofile2_5} are similar to the droplet profiles calculated in Refs.~\cite{hughes2014introduction,hughes2015liquid} using DFT for a very similar lattice-gas model; see also the 3D droplet density profile calculated using Monte Carlo simulations in Ref.\ \cite{chalmers2017modelling}.

\begin{figure}[t]
\includegraphics[width=20pc]{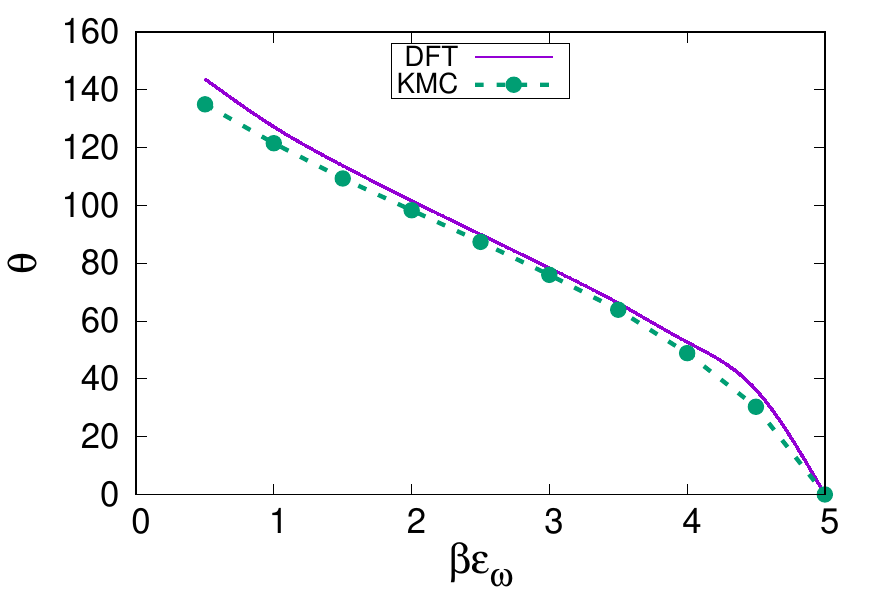}
\caption{\label{contactfig} The contact angle for droplets with temperature $\beta{\varepsilon} = 2.5$ on a surface with varying wall attraction strength $\varepsilon_w$. The solid line is the result obtained from the lattice DFT \eqref{eq:free_energy} together with Young's equation (\ref{eq:Young}). The dashed line is the results from KMC simulations.}
\end{figure}

From droplet density profiles such as those in Fig.\ \ref{fig:densityprofile2_5}, following e.g.\ Ref.~\cite{chalmers2017modelling}, we can calculate the contact angle by fitting a circle to the location of the liquid-vapour interface, which is defined as the points determined by linear interpolation between pairs of neighbouring lattice sites, where the density $\rho=0.5$. Results for the contact angle determined via this method are plotted in Fig.\ \ref{contactfig} for the temperature $\beta\varepsilon=2.5$ and for a range of different values of the wall attraction $\beta\varepsilon_w$. We also display the results from DFT and via Young's equation \eqref{eq:Young}. We see excellent agreement between the results from the two approaches, demonstrating that at such temperatures (away from the critical point) the DFT is accurate and so the binding potential obtained from the DFT should also be reliable.

The results in Fig.\ \ref{contactfig} show that the contact angle $\theta = 0$ for $\beta\varepsilon_w > 5$, where the liquid wets the wall. As $\varepsilon_w$ is decreased, decreasing the strength of the attraction between the particles and the surface so that the surface becomes more solvophobic, the contact angle $\theta$ increases and the liquid increasingly beads-up upon the surface.

 \section{Results for non-equilibrium droplets on surfaces}\label{sec:6}
 
  The results in the preceding sections essentially constitute a series of checks that the model has the desired, correct and self-consistent thermodynamics. In this section, we present results for the non-equilibrium behaviour predicted by the model. 

\subsection{Pure KMC at two different temperatures}

\begin{figure}
\includegraphics[width=5.9cm]{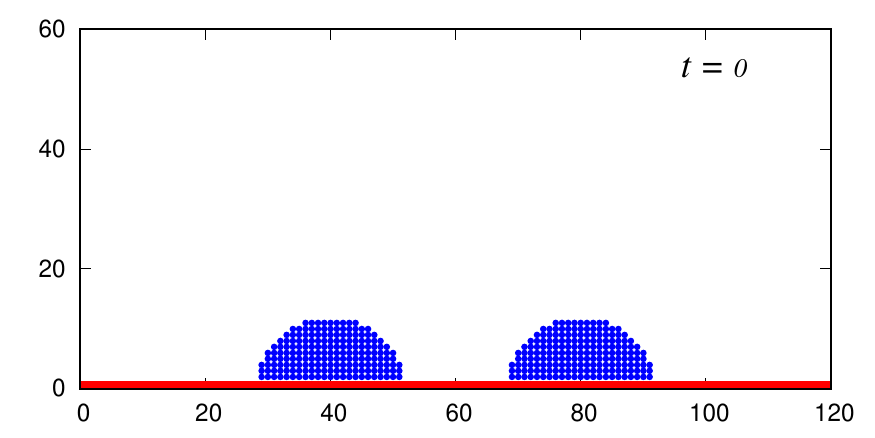}
\includegraphics[width=5.9cm]{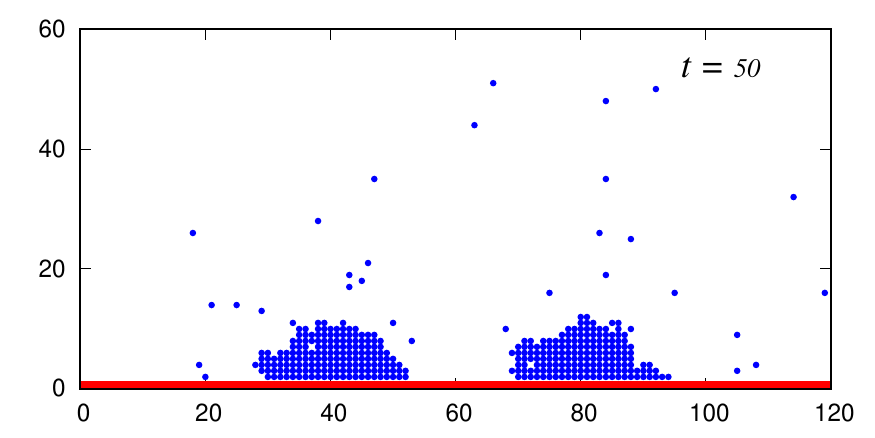}
\includegraphics[width=5.9cm]{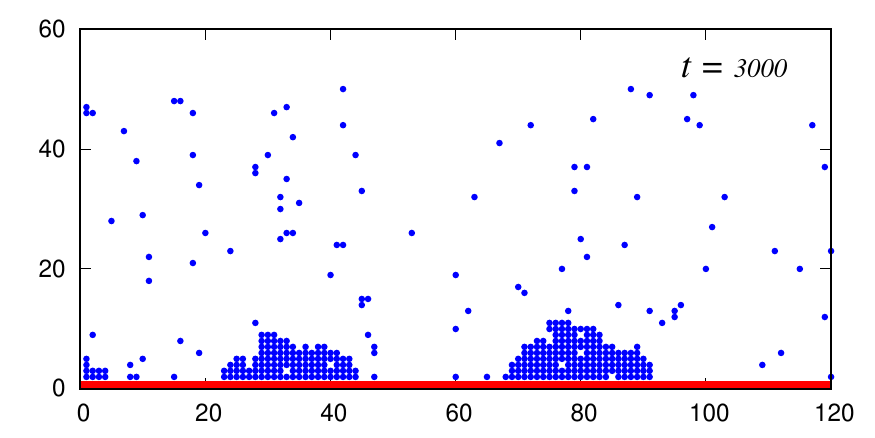}

\includegraphics[width=5.9cm]{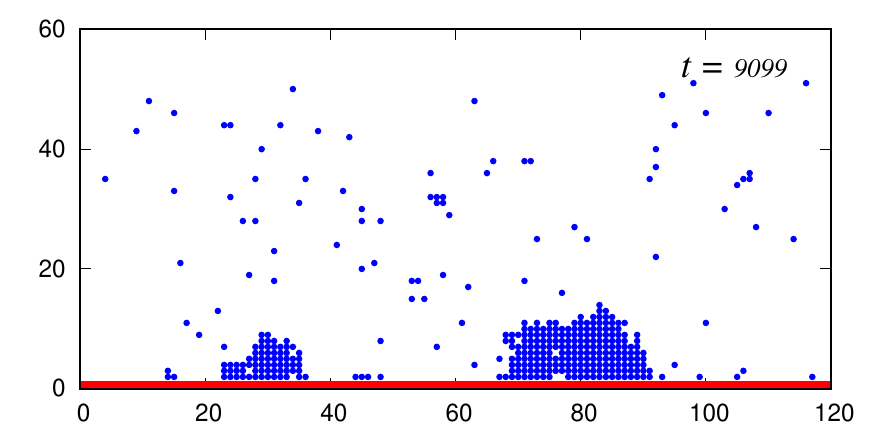}
\includegraphics[width=5.9cm]{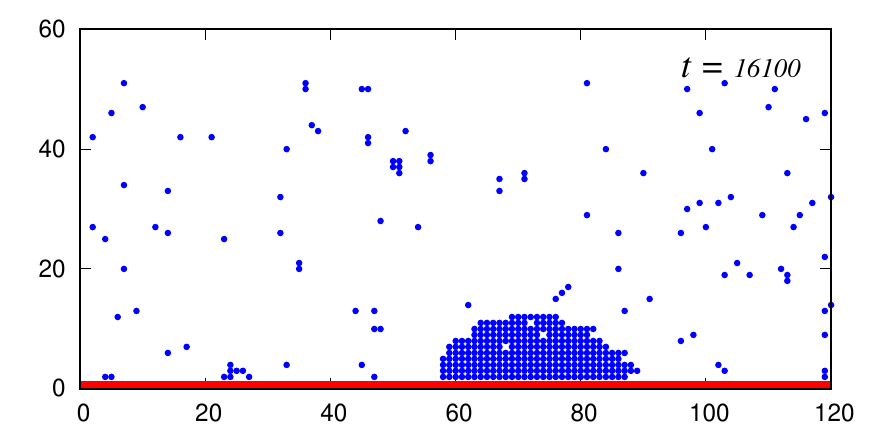}

\caption{Simulation of the coarsening (joining) of two droplets using KMC for $\beta\varepsilon=1.5$ and $\beta\varepsilon_{w}=2.1$ in a domain of length $120\sigma$ and height $60\sigma$. The times corresponding to each snapshot are in units of the KMC steps divided by $10^4$.}
\label{fig:PureKMC1_5}
 \end{figure}
 
\begin{figure}
\includegraphics[width=5.9cm]{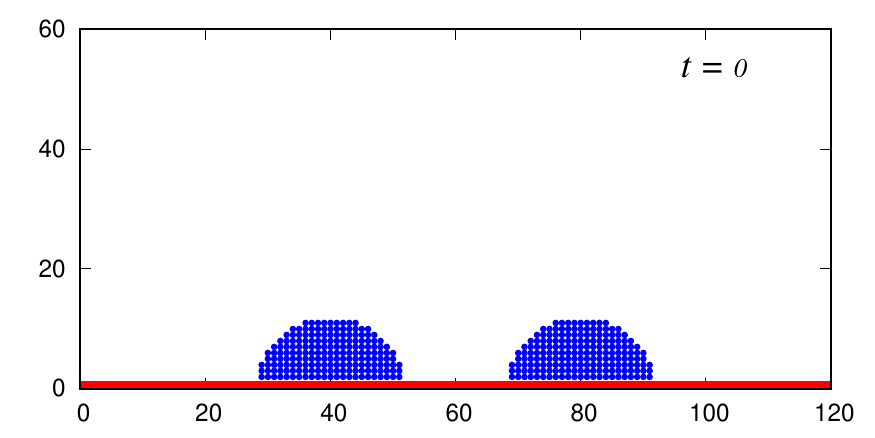}
\includegraphics[width=5.9cm]{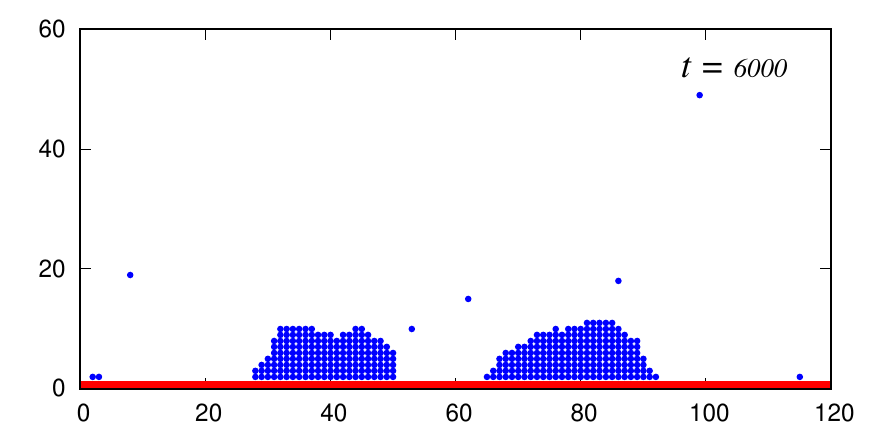}
\includegraphics[width=5.9cm]{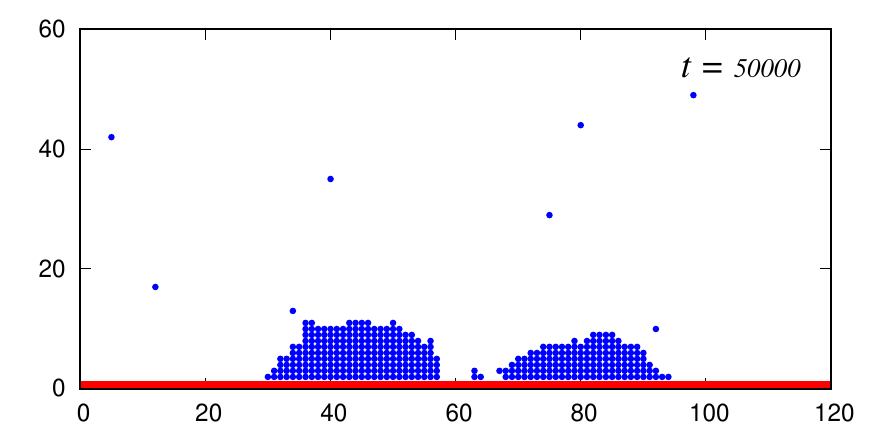}

\includegraphics[width=5.9cm]{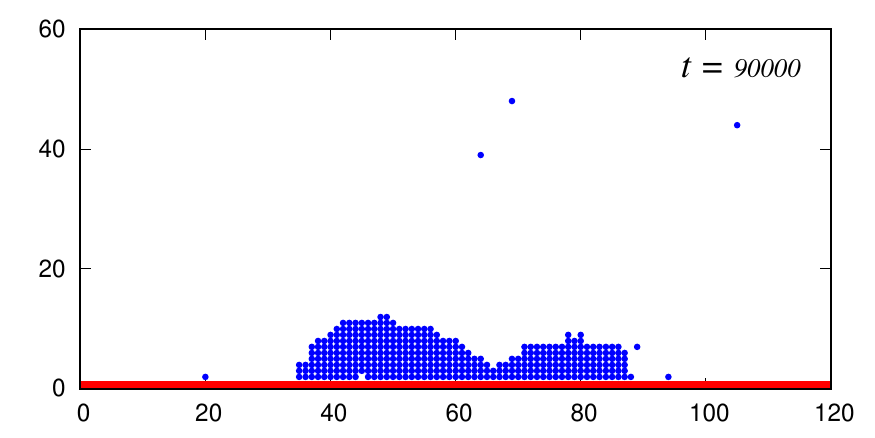}
\includegraphics[width=5.9cm]{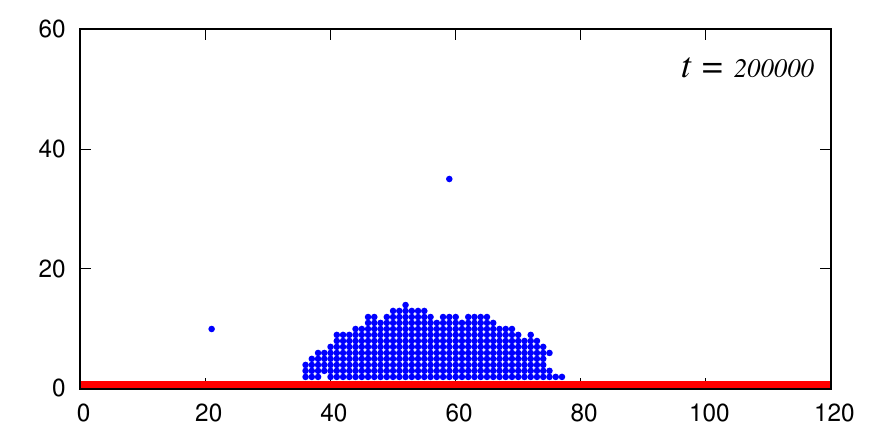}

\caption{Simulation of the joining (coarsening) of two droplets using KMC for $\beta\varepsilon=2.5$ and $\beta\varepsilon_{w}=3.5$ in a domain of length $120\sigma$ and height $60\sigma$. The times corresponding to each snapshot are in the units of KMC steps divided by $10^4$.}
\label{fig:PureKMC2_5}
\end{figure}

 For droplets to be able to equilibrate with the vapour phase, the system must be enclosed (i.e.\ treated in the canonical ensemble), with a fixed number of particles in the system. This is achieved by assuming a hard but non-attracting ($\varepsilon_w=0$) wall on the top.	

We first present results for the behaviour of two droplets using only KMC simulations. The total domain is of size $120\sigma\times 60\sigma$, with periodic boundary conditions between the left and right sides. The system is initiated by setting $l_\bold{i}=1$ for all the lattice sites within two semi-circular regions of radii $12\sigma$ with the centres of the corresponding circles at the sites $(40,1)$ and $(80,1)$, and we set $l_\bold{i}=0$ outside of these two semi-circles. This gives $N=368$ particles in the system.

The results in Fig.~\ref{fig:PureKMC1_5} are for the temperature $k_{B}T/\varepsilon = 1/1.5$ and we choose the wall attraction parameter $\beta\varepsilon_{w}=2.1$. The contact angle for these parameter values is $\theta \approx 60^\circ$. The different panels in Fig.~\ref{fig:PureKMC1_5} correspond to a sequence in time, i.e.\ for increasing numbers of the KMC steps, as indicated in the top right of each. The droplets initially evaporate and somewhat decrease in size, equilibrating with a low density vapour phase. They also adjust their average contact angle from the initial value of $\theta = 90^\circ$ closer to $\theta \approx 60^\circ$, the equilibrium value (c.f.~Fig.~\ref{fig:densityprofile2_5}). However, as time proceeds, due to the random (thermal) fluctuations, the size of one of the droplets (the left one in this KMC simulation) keeps decreasing, but the size of the other droplet (the right one in this KMC simulation) starts to increase. The centres of the droplets remain at approximately the initial positions. Eventually, the left droplet completely disappears, leaving the system containing only a single droplet, which of course is the global free energy minimum state. In the case in Fig.~\ref{fig:PureKMC1_5} (i.e.\ for the temperature $k_{B}T/\varepsilon = 1/1.5$), this coarsening process occurs mainly via the diffusive migration of the particles through the vapour, evaporating from one droplet and condensing onto the other, although there is additionally a small particle flux over the surface. In Ref.~\cite{pototsky2014coarsening}, when clusters (equivalent to our droplets) aggregate in a similar manner, it is referred to as joining via the `Ostwald mode', in contrast to joining by the droplets moving towards one another, which is referred to as joining via the `translation mode'. For the model system considered in Ref.~\cite{pototsky2014coarsening}, it is possible to show that there is an eigenfunction associated with each of these two possible modes and that it is the mode with the largest eigenvalue that dominates the observed dynamics, although in principle it is possible for the two eigenvalues to be similar in value. However, in the case in Fig.~\ref{fig:PureKMC1_5} it is clearly the Ostwald mode that is dominant. For more background on Ostwald ripening, see e.g.\ Ref.~\cite{ratke2013growth}.

Figure\ \ref{fig:PureKMC2_5} correspond to a lower temperature $k_{B}T/\varepsilon = 0.4$ and we choose the wall attraction parameter $\beta\varepsilon_{w}=3.5$ so that the equilibrium contact angle is still $\theta \approx 60^\circ$. The droplets in this case initially spread a little over the surface because the KMC time evolution drives the system towards achieving the equilibrium contact angle for both droplets. There is a much smaller rate of evaporation because of the lower temperature, or, equivalently, due to the stronger attraction between the particles in this case. Then over time the droplets eventually merge and form a single droplet in the centre of the domain.  The coarsening process in this case appears to mainly be due to overall translation of the droplets over the surface, i.e. coarsening due to the translational mode \cite{pototsky2014coarsening,ratke2013growth}.


 \subsection{Combination of KMC and thin-film equation dynamics}

\begin{figure}
\includegraphics[width=5.9cm]{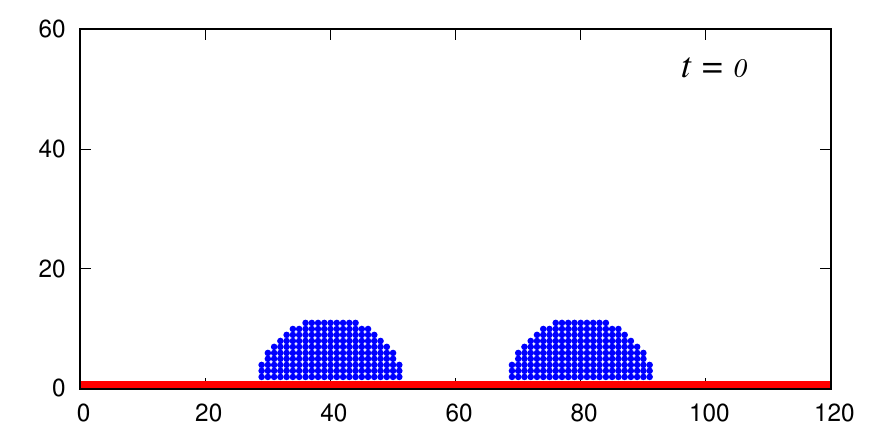}
\includegraphics[width=5.9cm]{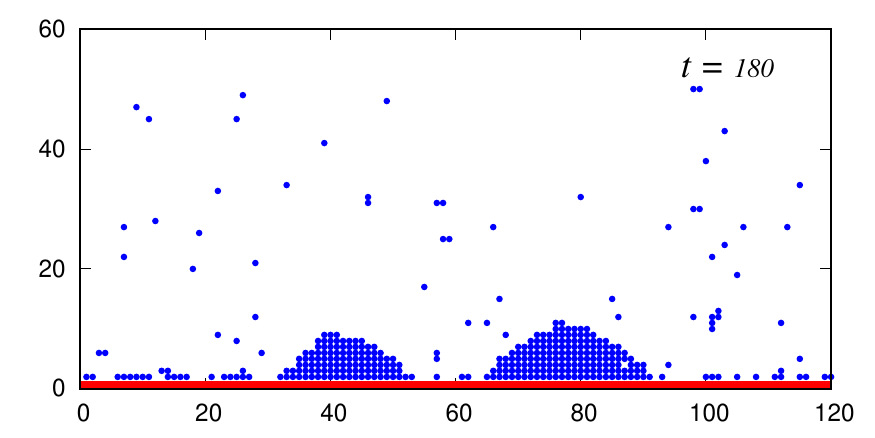}
\includegraphics[width=5.9cm]{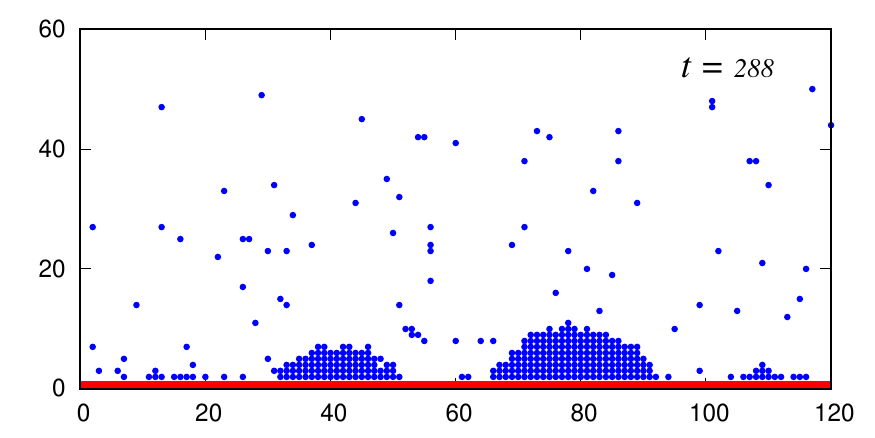}

\includegraphics[width=5.9cm]{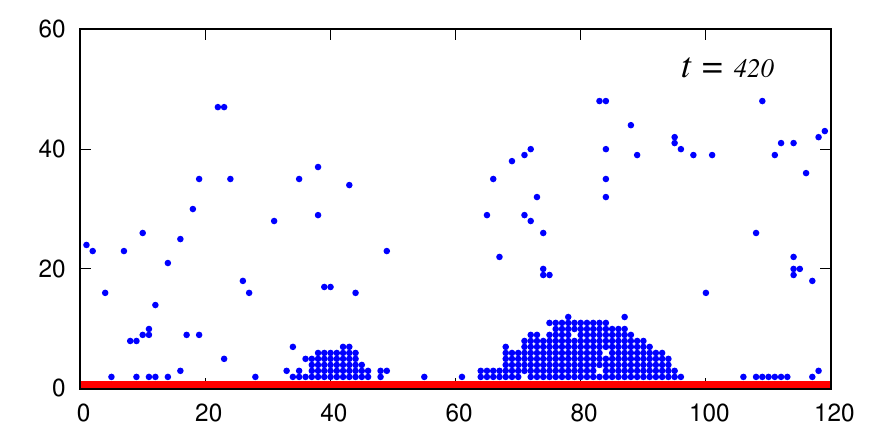}
\includegraphics[width=5.9cm]{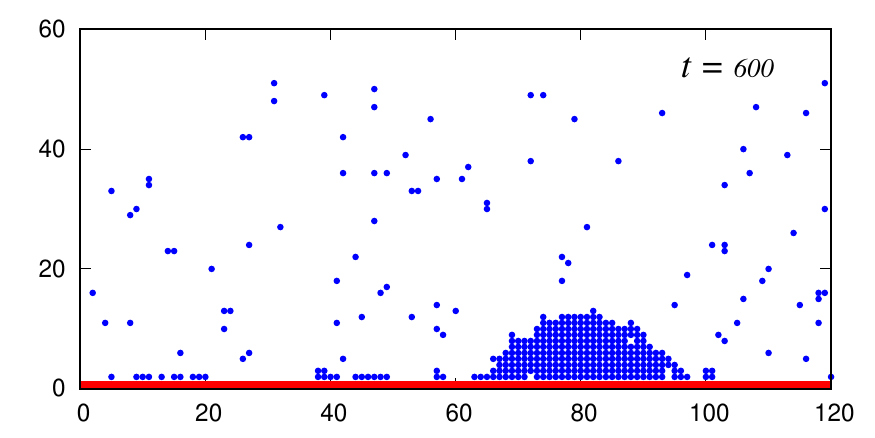}
\includegraphics[width=5.9cm]{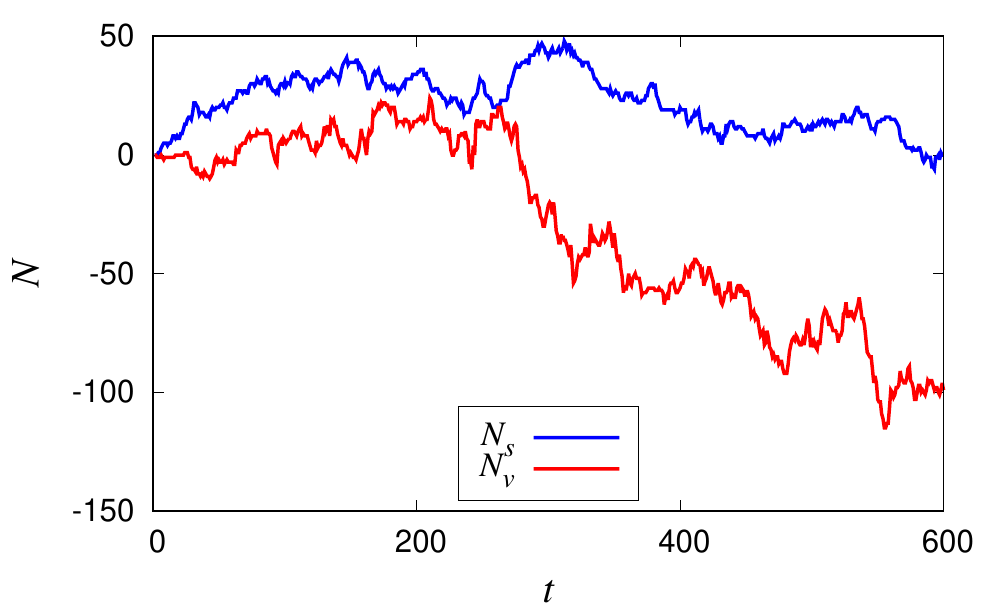}
\caption{Simulation of two droplets joining, with the flux from one to the other being largely via diffusion through the vapour. These results are for $\beta\varepsilon=1.5$ and $\beta\varepsilon_{w}=2.1$ in a domain of size $120\sigma\times60\sigma$. We calculate over a total of $n=600$ cycles, with each cycle consisting of $M_1=10^4$ KMC steps alternating with $M_2=10^4$ thin-film steps with $\Delta t=10^{-5}$. The times indicated on each snapshot correspond to the number of cycles passed. In the final panel we show the time evolution of the numbers of the particles that cross the vertical line in the middle of the domain, at $i=60$, migrating over the surface, $N_s$, (blue line) and through the vapour, $N_v$, (red line). Increasing/decreasing $N_s$ or $N_v$ means that the particle moves to the left/right, respectively. Since $|N_v|\gg|N_s|$, this indicates that diffusion through the vapour is the dominant process.}
\label{fig:KMC&thin1_5} 
\end{figure}

\begin{figure}
\includegraphics[width=5.9cm]{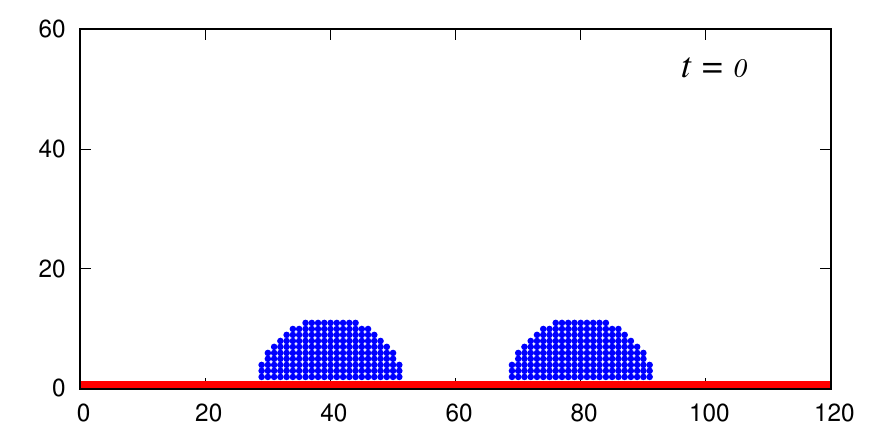}
\includegraphics[width=5.9cm]{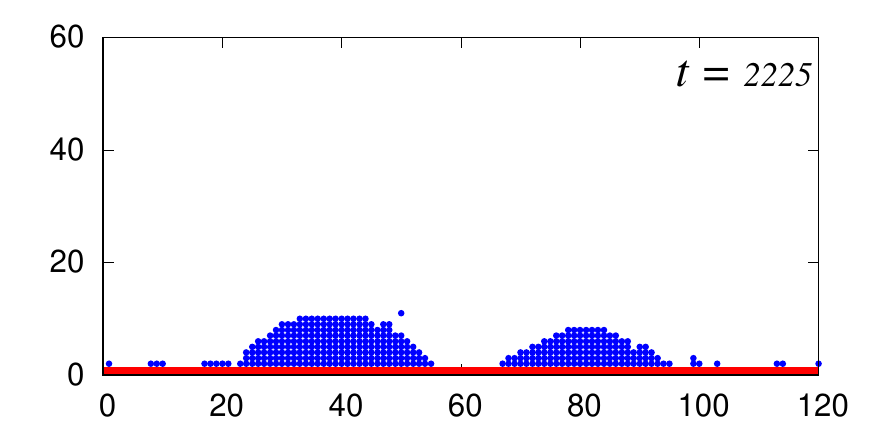}
\includegraphics[width=5.9cm]{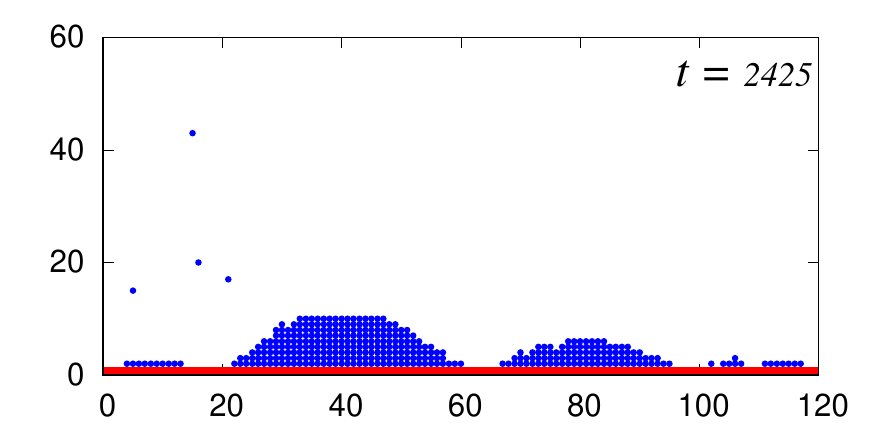}

\includegraphics[width=5.9cm]{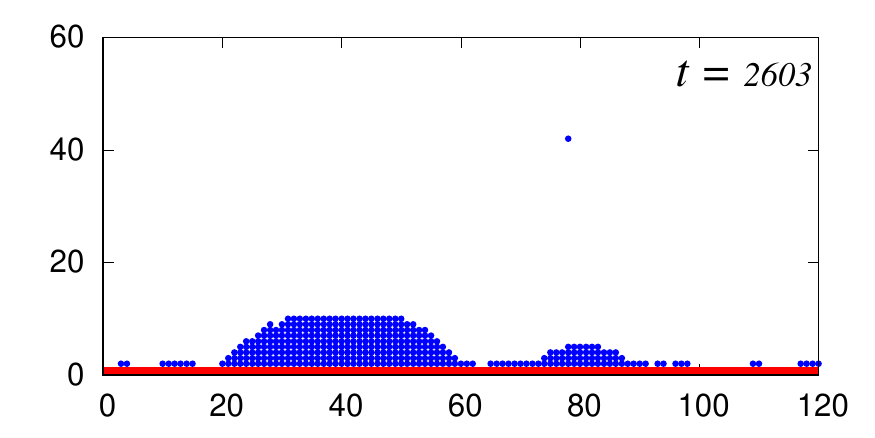}
\includegraphics[width=5.9cm]{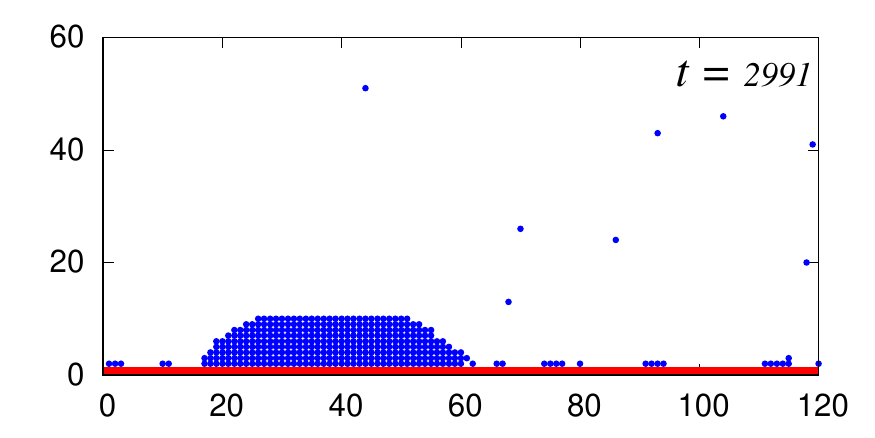}
\includegraphics[width=5.9cm]{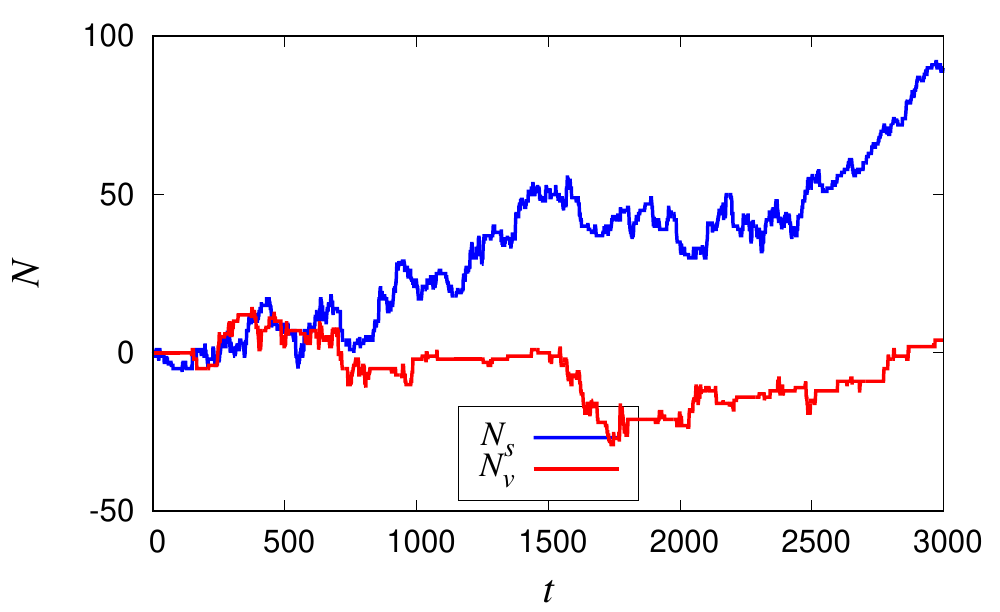}
\caption{Simulation of two droplets joining, with time evolution including both single-particle KMC dynamics and also thin-film equation dynamics, for $\beta\varepsilon=2.5$ and $\beta\varepsilon_{w}=3.5$ in a domain of size $120\sigma\times60\sigma$. We have a total of $n=3000$ cycles consisting of $M_1=10^4$ KMC steps and $M_2=10^4$ thin-film steps with $\Delta t=10^{-5}$. The times corresponding to each snapshot are in the units of cycles. The last panel displays the time evolution of the numbers of particles that cross the vertical line in the middle of the domain at $i=60$, migrating over the surface, $N_s$, (blue line) and through the vapour, $N_v$, (red line). Increasing/decreasing $N_s$ or $N_v$ means that the particle moves to the left/right, respectively.}
\label{fig:KMC&thin2_5}  
\end{figure} 

We now consider behaviour of the system when the fluid dynamics governed by a combination of the single-particle KMC moves in conjunction with the collective particle moves generated by the thin-film equation. We use the same domain and the same initial configuration as in the previous subsection. Figure \ref{fig:KMC&thin1_5} corresponds to the same temperature and wall attraction strength as in Fig.~\ref{fig:PureKMC1_5}. For this simulation, we used $n=600$ cycles of $M_1=10^4$ KMC steps followed by $M_2=10^4$ thin-film steps with the time step $\Delta t=10^{-5}$, which gives $\lambda=272$, where $\lambda$ is the parameter defined by Eq.\ (\ref{eq:lambda}). As in Fig.\ \ref{fig:PureKMC1_5}, the system coarsens into one droplet. However, the speed of coarsening is increased when the thin-film steps are included, which is not too surprising, since the total number of possible moves has been increased by the addition of the thin-film moves. As before, in the initial stages the droplets partially evaporate and spread, to reach a contact angle close to the equilibrium value. In the later stages, it can be observed that the size of one of the droplets (the left one in this simulation) keeps decreasing, but the size of the other droplet (the right one in this simulation) starts to increase. As in the case in Fig.~\ref{fig:PureKMC1_5}, the centres of the droplets in Fig.~\ref{fig:KMC&thin1_5} remain more or less fixed. The coarsening in this case is mainly due to a flux of particles evaporating from one droplet, travelling though the vapour and condensing onto the other, although there is also a smaller flux due to a diffusive migration of the particles over the surface. To illustrate this in more detail, we show in the bottom right panel of Fig.\ \ref{fig:KMC&thin1_5} how the numbers of particles that cross the vertical line in the middle of the domain at $i=60$ varies in time. These are split into two populations: (i) those that migrate over the surface at height $j\leq h_i$ ($N_s$, see the blue line) and (ii) those that travel through the vapour at height $j>h_i$ ($N_v$, see the red line). Increasing/decreasing $N_s$ or $N_v$ means that the particle moves to the left/right, respectively, crossing the $i=60$ line. In this particular simulation, it can be observed that there is relatively low migration of the particles over the surface, whilst there is a significant net transfer of particles from left to right through the vapour. Thus, we can conclude that in this case the main coarsening mechanism is due to particles evaporating from one droplet, diffusing through the vapour, and then condensing onto the other droplet. Using only the thin-film equation would not allow us to capture this, and this justifies the importance of our hybrid model which combines both the thin-film equation governed dynamics with the KMC diffusive dynamics. 

Figure\ \ref{fig:KMC&thin2_5} corresponds to the same values of the temperature and the wall attraction strength parameter as in Fig.~\ref{fig:PureKMC2_5}. For this simulation, we used $n=3000$ cycles of $M_1=10^4$ KMC steps followed by $M_2=10^4$ thin-film steps with the time step $\Delta t=10^{-5}$, which gives the same value of $\lambda$ as for the results in Fig.\ \ref{fig:KMC&thin1_5}. As in the higher temperature case in Fig.~\ref{fig:KMC&thin1_5}, we find that the speed of the coarsening is increased when the thin-film steps are included. As in the pure KMC case in Fig.\ \ref{fig:PureKMC2_5}, the droplets initially spread over the surface to reach a contact angle close to the equilibrium value $\theta \approx 60^\circ$, but with only a small amount of evaporation due to the low temperature. As time proceeds, the size of the right droplet starts to decrease and the left droplet starts to grow, and apparently this happens mainly due to motion of particles from one to the other over the surface. To understand the coarsening mechanisms in more detail, we present in the last panel of Fig.\ \ref {fig:KMC&thin2_5} the time evolution of the numbers of the particles that cross the vertical line in the middle of the domain, at $i=60$, that migrate over the surface, $N_s$,  and through the vapour $N_v$. In this particular simulation, it can be observed that there is a relatively low rate of migration of the particles through the vapour, whilst the majority of the mass transfer is via the surface. Thus, we can conclude that in this case of lower temperature, the main coarsening mechanism is due to the transfer of the particles over the surface.


\subsection{Droplet evolving under lateral driving}\label{subsec:VIC}

\begin{figure}
\includegraphics[width=5.9cm]{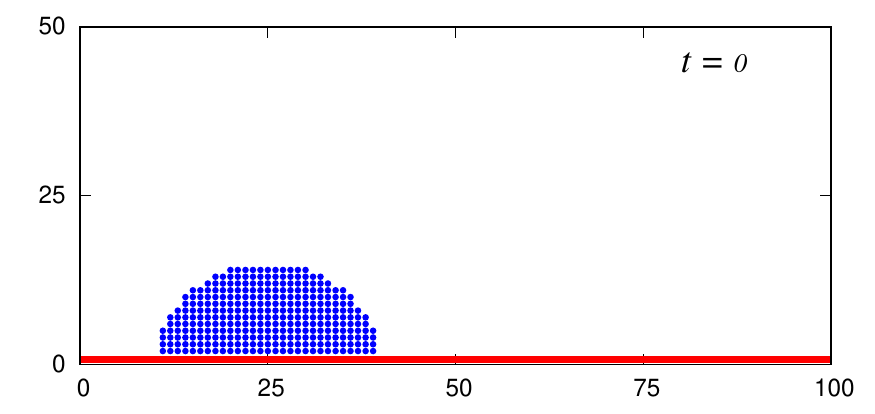}
\includegraphics[width=5.9cm]{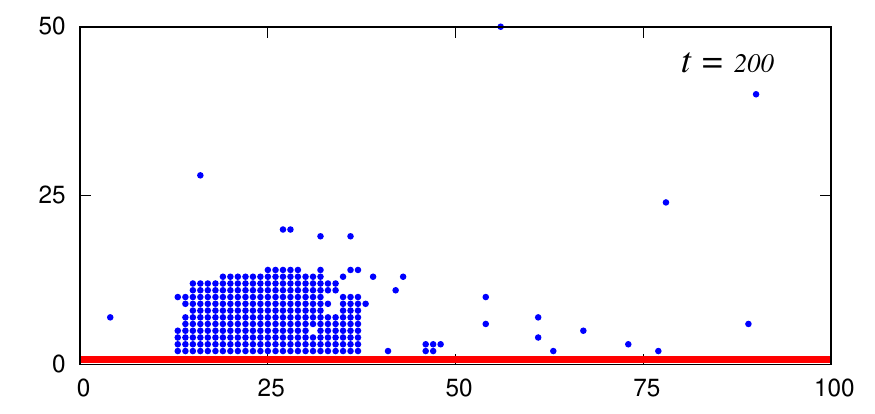}
\includegraphics[width=5.9cm]{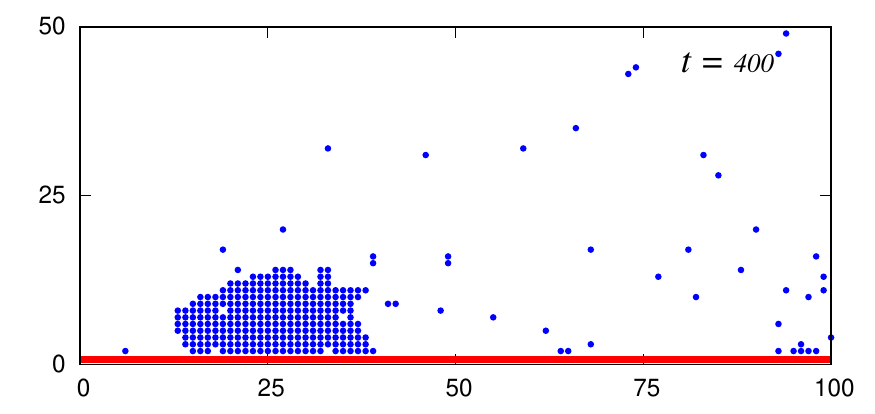}

\includegraphics[width=5.9cm]{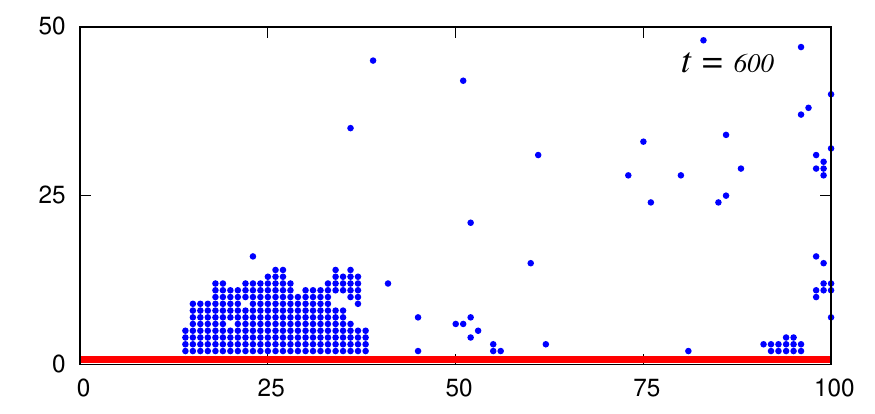}
\includegraphics[width=5.9cm]{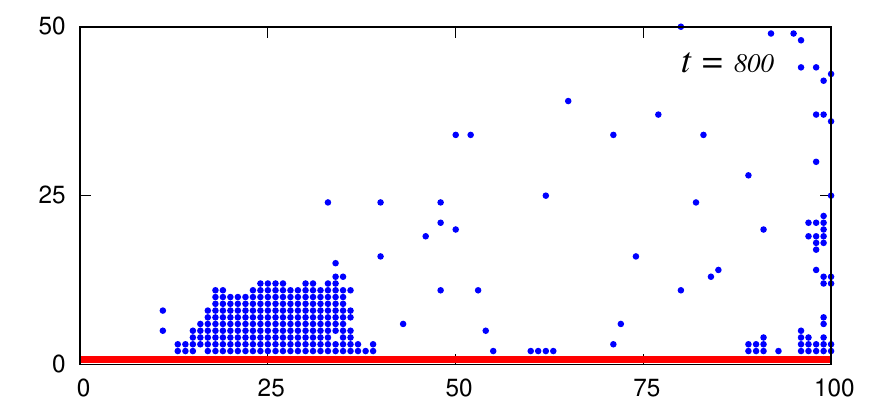}
\includegraphics[width=5.9cm]{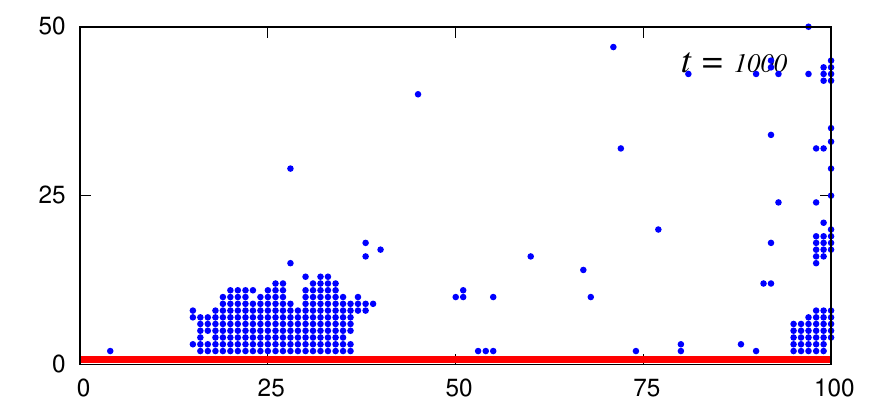}

\caption{A droplet evolving in time under the influence of a constant force ${\cal G} = 0.2$ to the right, for $\beta\varepsilon=1.5$ and $\beta\varepsilon_{w}=2.1$. The dynamics is pure single-particle KMC (i.e.\ $\lambda=\infty$) and the system is of size $100\sigma\times50\sigma$. The times corresponding to each snapshot are in units of KMC steps divided by $1500$, (i.e.\ $M_1=1500$ and $M_2=0$).}
    \label{fig:pure_KMC_force}
\end{figure}

 \begin{figure}
\includegraphics[width=5.9cm]{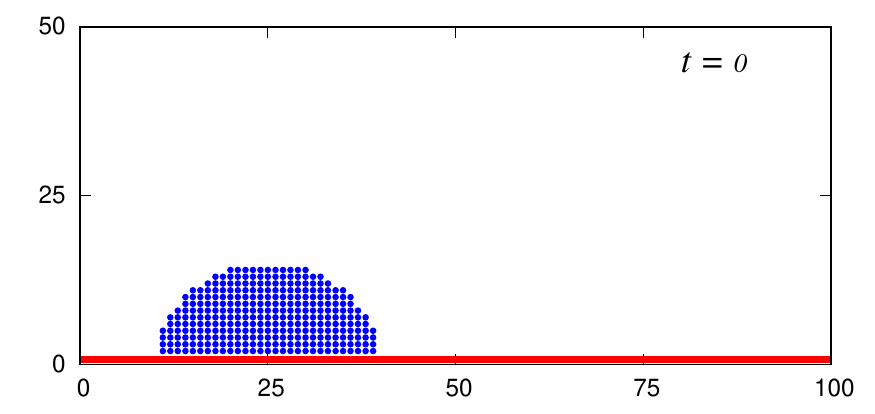}
\includegraphics[width=5.9cm]{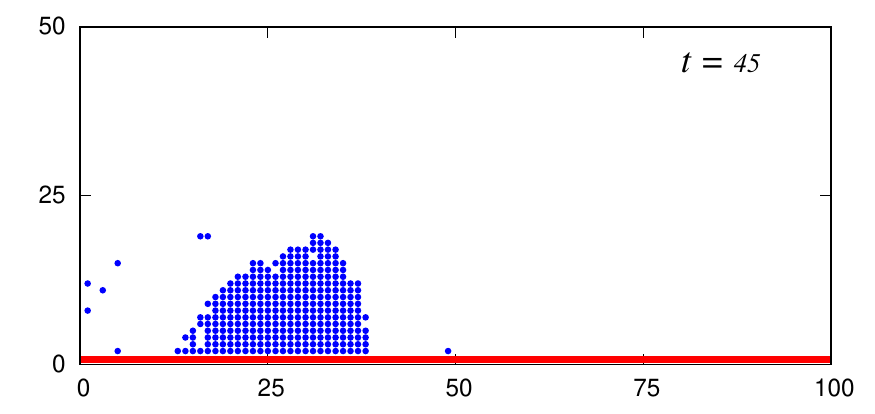}
\includegraphics[width=5.9cm]{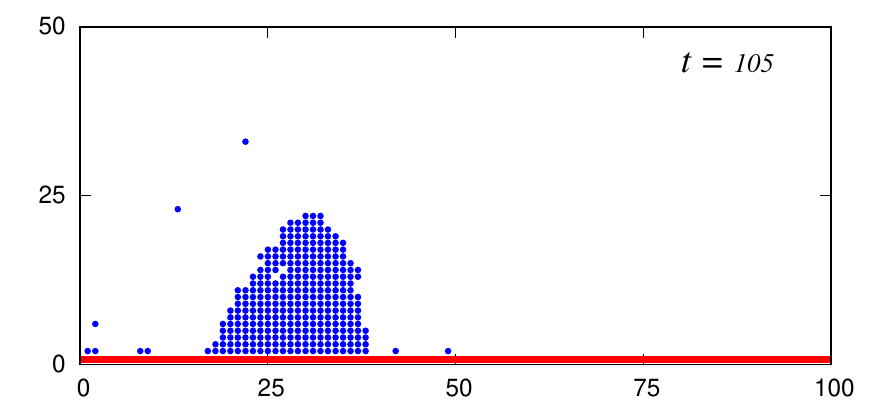}

\includegraphics[width=5.9cm]{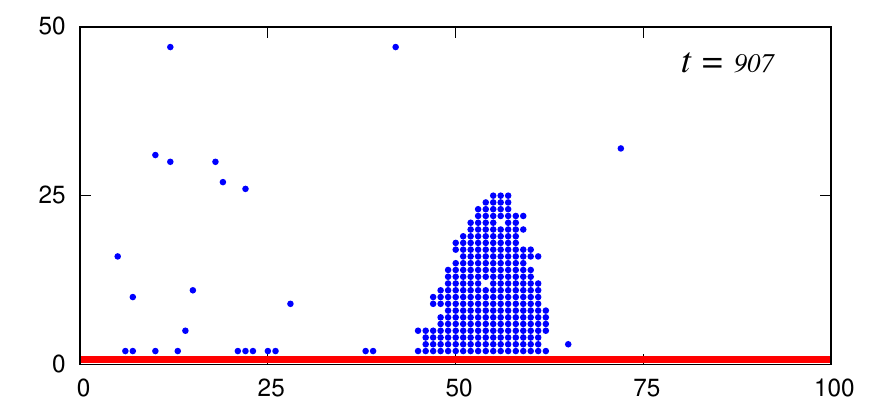}
\includegraphics[width=5.9cm]{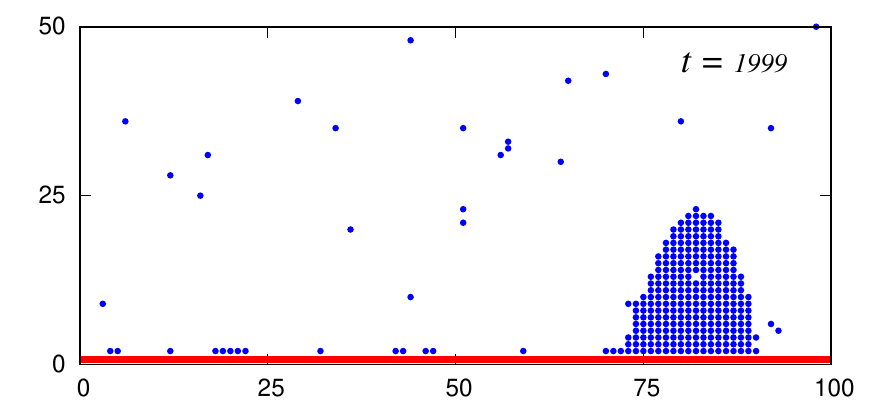}
\includegraphics[width=5.9cm]{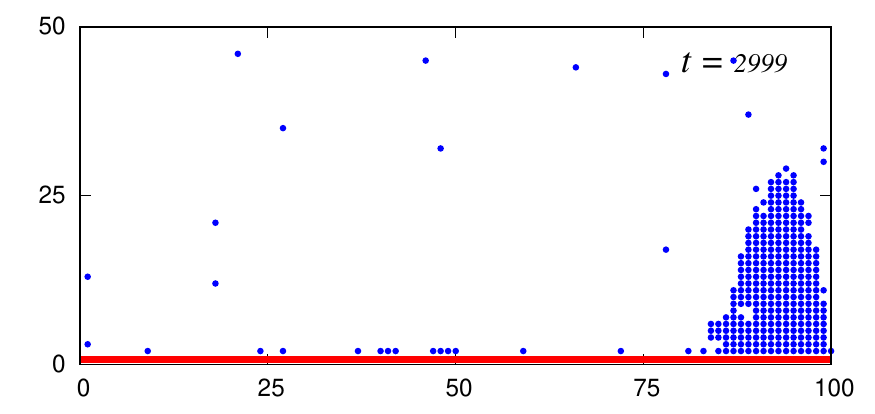}

\caption{Results for a system the same as that in Fig.~\ref{fig:pure_KMC_force}, except here the particle motion also includes collective multi-particle moves generated by the thin-film equation, as well as some single particle moves from the KMC. We have $M_1=1.5\times10^3$ KMC steps per cycle and $M_2=10^3$ thin-film steps with $\Delta t=10^{-6}$ per cycle, which corresponds to $\lambda=4918$. The times corresponding to each snapshot are in the units of numbers of cycles.}
\label{fig:force1}
\end{figure}

\begin{figure}
\includegraphics[width=5.9cm]{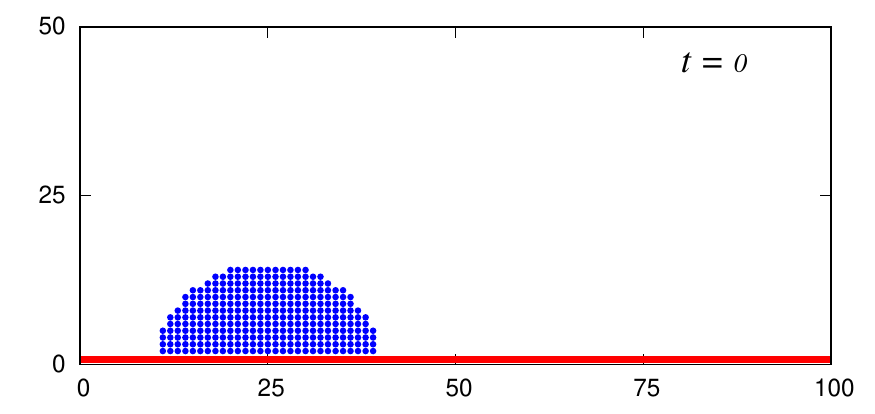}
\includegraphics[width=5.9cm]{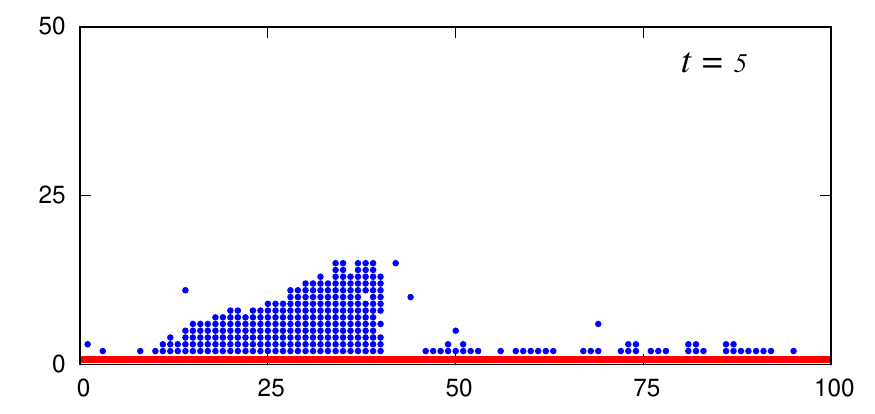}
\includegraphics[width=5.9cm]{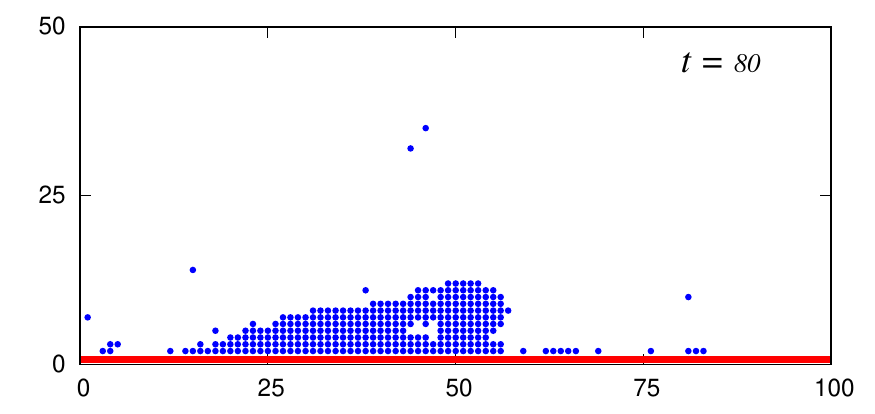}

\includegraphics[width=5.9cm]{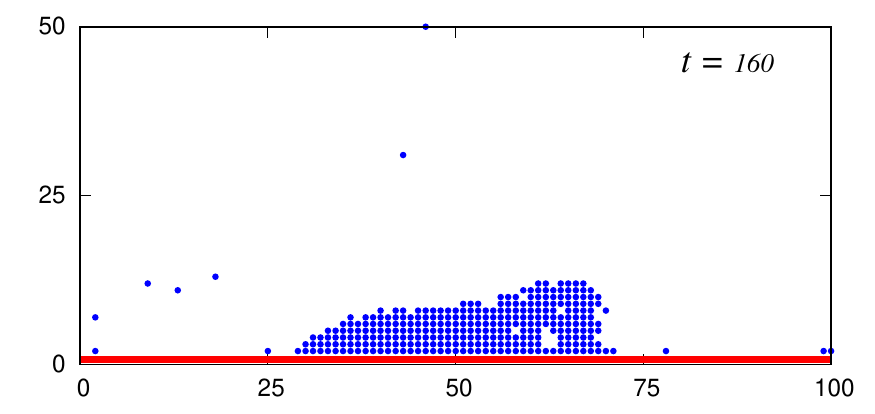}
\includegraphics[width=5.9cm]{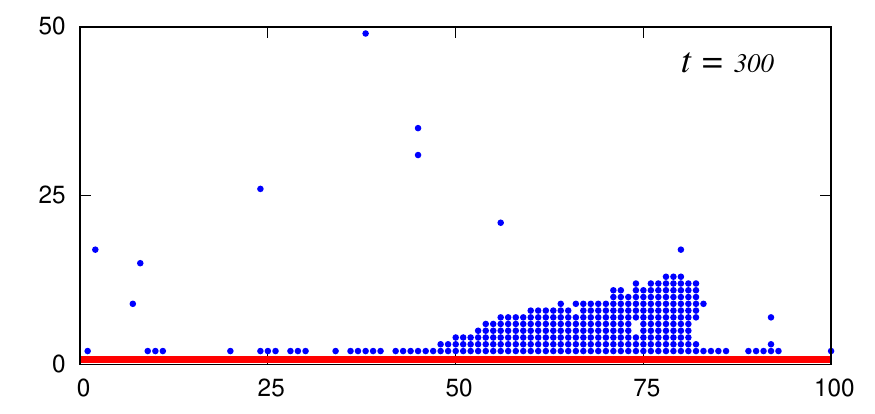}
\includegraphics[width=5.9cm]{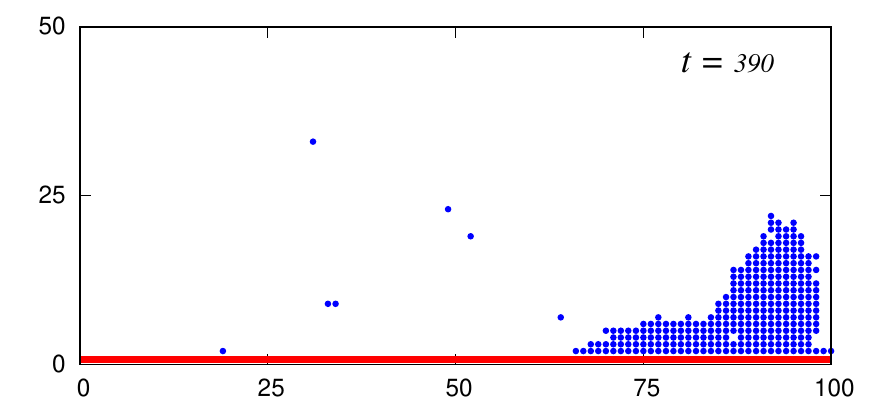}
\caption{Results for a system the same as that in Figs.~\ref{fig:pure_KMC_force} and \ref{fig:force1}, except here there are an even higher proportion of the dynamics constituting of collective multi-particle moves generated by the thin-film equation. We have $M_1=1.5\times10^3$ KMC steps per cycle and $M_2=10^5$ thin-film steps with $\Delta t=10^{-6}$ per cycle, which corresponds to $\lambda=0.49$. The times corresponding to each snapshot are in the units of numbers of cycles.}
\label{fig:force2}
\end{figure}

In this subsection, we consider the dynamics predicted by the model when the fluid is under the influence of a constant lateral driving force parallel to the surface, i.e.\ we have a constant force ${\cal G}=0.2$ to the right on each particle. The domain is a box of size $100\sigma\times50\sigma$, with hard walls on both the left and right boundaries, as well as the attracting surface on the bottom and the hard wall at the top considered previously. The system is initiated by setting $l_\bold{i}=1$ for all the lattice sites within a semi-circular region of radius $14\sigma$ with the centre at the site $(25,1)$, and we set $l_\bold{i}=0$ outside of this semi-circle. This gives $N=305$ particles in the system.  As a consequence of the lateral driving and the walls at the left and right boundaries, in all of the results displayed in this section in the final stages of the dynamics we see a pile-up of particles on the right-hand boundary wall. 

Figure \ref{fig:pure_KMC_force} shows results for when the dynamics is solely generated by the KMC algorithm, with no thin-film dynamics. The initial configuration consists of the semi-circular droplet displayed in the top left panel, which has contact angle $\theta\approx60^\circ$. We set $\beta\varepsilon = 1.5$ and the wall attraction strength $\beta\varepsilon_{w}=2.1$. We see from the results in Fig.~\ref{fig:pure_KMC_force} that the droplet centre of mass hardly moves over time. However, since there is a constant (but slow) evaporation rate of particles from the surface, once these leave the droplet they are then carried away to the right by the driving force ${\cal G}\neq0$. Over time, these then build up on the wall on the right side. Thus, although on every particle in the system there is a force to the right ${\cal G}=0.2$, due to the KMC dynamics only considering one particle at a time, we see that there is little or no collective motion resulting from this force. As we see next, this is not the case when we also allow collective motion by considering the dynamics from the thin-film equation.

Figure \ref{fig:force1} shows snapshots from a simulation where all the model parameters are the same as for the simulation in Fig.\ \ref{fig:pure_KMC_force} except we now include some collective multi-particle moves generated by considering the thin-film equation \eqref{eq:thin} with ${\cal G}=0.2$. We set $M_1=1.5\times10^3$, $M_2 = 10^3$ and $\Delta t = 10^{-6}$, which gives $\lambda= 4918$. We see that the thin-film moves enable the droplet to migrate across the surface under the lateral driving. By comparing with the previous result in Fig.~\ref{fig:pure_KMC_force}, it is clear that this dynamics must be a collective multi-particle (hydrodynamic) phenomenon. Though the majority of the particle motion to the right is now via the droplet sliding over the surface, there is still a small flux through the vapour. Note too that the overall timescale for the dynamics is reduced, due to the additional thin-film moves.

In Fig.~\ref{fig:force2} we show results for a system with an even higher proportion of particle moves generated by the thin-film equation. In this case we have $M_1=1.5\times10^3$, $M_2 = 10^5$ and $\Delta t = 10^{-6}$, which gives $\lambda=0.49$. We again observe that the thin-film moves enable the droplet to slide from left to right under the driving. Additionally, for this choice of parameter values we see that the droplet takes a characteristic shape with a `tail' trailing behind, typical sometimes of rain-droplets sliding down a window. The appearance and bifurcations of such tails is discussed in Ref.~\cite{wilczek2017sliding} for the thin-film equation.


\section{Concluding remarks}\label{sec:7}

We have developed a hybrid model for droplet dynamics on surfaces, that combines both single-particle diffusive dynamics via KMC and advective hydrodynamic multi-particle moves via a thin-film equation. We have shown that the model incorporates the correct thermodynamics and hydrodynamics, microscopic structure, evaporation, condensation and the diffusion of particles over a surface. The model contains parameters which can be determined by relating to experiments. For example, from measurements of the equilibrium contact angle of drops of the liquid on the relevant surfaces and the surface tension, the interaction parameters can be estimated and the mobility coefficients estimated from the bulk liquid viscosity and diffusion coefficient -- see also the discussion in Ref.~\cite{chalmers2017modelling} about determining these coefficients.

An important part of the coarse-graining that we performed in order to bridge from the microscopic Hamiltonian \eqref{eq:Hamiltonian} to the mesoscopic thin-film equation \eqref{eq:thin}, is the use of DFT. Here we considered a lattice model, to make this bridging more easily. However, we see no reason why this could not also be done for a continuum model. The main requirement is to have an accurate DFT, that is able to describe the structure of the liquid of interest. For example, for the Lenard-Jones model liquid, much of the work has already been done in Ref.~\cite{hughes2017}. The one thing that would need to be adapted and developed further is the algorithm for determining the liquid droplet height profile and the algorithm for moving the particles as determined by the thin-film equation. For the former, we believe that calculating a local adsorption would be sufficient. However, the present algorithm for inserting the particles would surely need some testing and refining, since in the present work the underlying lattice makes this process much more straight-forward to implement. We should also emphasise how valuable it is to make the connection with DFT, since this gives access to thermodynamic quantities such as the surface tensions, equilibrium contact angles, pressure and the bulk fluid phase behaviour. These are important things to know and are much harder to obtain via other methods; for example, to obtain these from KMC simulations is more complicated and requires more lengthy computations.

For the purposes of the present study the DFT results presented here in Sec.~\ref{sec:DFT} were performed in order to make sure the thermodynamic underpinnings of the thin-film equation used in the time evolution is correct. However, we believe the present work is also a valuable contribution to the development and testing of lattice-DFT models of this type. There has previously been some testing of the mean-field DFT \eqref{eq:free_energy} by comparing to simulations -- see e.g.\ Ref.~\cite{chalmers2017dynamical} -- but the results presented here showing that the DFT is also accurate at some temperatures for describing the contact angle and the other related interfacial properties have not been performed before for a 2D model, to the best of our knowledge.

We have presented striking results showing the process of how a pair of droplets on a surface join together depends on the temperature and also the other system parameters: at higher temperatures the coarsening process largely occurs via evaporation of particles from the smaller droplets, diffusion through the vapour and then condensation onto the larger droplets, i.e.\ via the Ostwald mode \cite{pototsky2014coarsening}. In contrast, at lower temperatures drops coalesce via the translation mode, moving towards one another over the surface and joining. In the present lattice system it is not possible to calculate the eigenfunctions and eigenvalues associated with each of these modes, as it is for the simpler continuum model in Ref.~\cite{pototsky2014coarsening}. However, it is clear that these processes are present here. In fact, there are tantalising hints in some of our results that there may in fact be two different surface modes: an Ostwald-type mode, determined by the diffusion of particles over the surface (rather than via the vapour), which is somewhat different in character to the collective surface motion due to capillary forces that is described by the thin-film equation. However, to confirm this conjecture and really distinguish that there are two distinct surface modes would require a much simpler continuum model that is amenable to the analysis approach used in Ref.~\cite{pototsky2014coarsening}.

The results presented in Sec.~\ref{subsec:VIC}, where the system evolves under lateral driving, show that collective many-particle hydrodynamic motion is essential for droplets to slide over surfaces. The thin-film equation dynamics is also crucial to whether the droplet takes a sliding shape with a tail or not. If the particles are assumed to move one at a time, then the droplets do not slide. The knowledge that the overall droplet shape is linked to the underlying particle motion gives valuable insight into the behaviour of these types of systems.

{For the driven droplets in Fig.~\ref{fig:pure_KMC_force}, where the dynamics is solely due to the KMC ($\lambda=\infty$), we see that the advancing and receding contact angles appear to be the same as the equilibrium contact angles. However, when $\lambda$ is finite (Figs.~\ref{fig:force1} and \ref{fig:force2}), clearly the advancing and receding contact angles are different. It would be of interest to determine how these angles depend on $\lambda$ and on the strength of the driving ${\cal G}$ and to compare with predictions in the literature, such as Refs.~\cite{BH69,V76}. However, to get the same amount of data in such non-equilibrium situations as was used to determine the equilibrium contact angle values with the statistical precision displayed in Fig.~\ref{contactfig} would require numerous repeated simulations. It is a much more complex scenario than the equilibrium case and so would require a very careful analysis of the results to match together all the different simulations, for robust conclusions to be obtained. This would enable the calculation of the non-equilibrium ensemble-average density profiles (the analogue of Fig.~\ref{fig:densityprofile2_5}), which would yield further information about the complex relationship between the contact line structure (density profile) and the wetting dynamics \cite{Wang}. Therefore, we leave this for future work.

One limitation of our model, inherited from using the thin-film equation \eqref{eq:thin}, is that for contact angles $\theta\geq90^\circ$ the model becomes inaccurate. The KMC part of the model works fine in this regime, it is only the thin-film equation part that fails. Thus, for example, the model could not be applied in its present form to study droplets on superhydrophobic surfaces. A further limitation of the model is that in certain situations it does not correctly capture the fluid dynamics within the droplet. For example, sometimes there can be a circulation of the liquid within a droplet as it slides \cite{Thampi2013,Karapetsas2016}. Some aspects of circulatory motion is described by the thin film equation, see e.g.\ \cite{Dmitri}. However, in our model if one were to tag a particular lattice particle, one would not observe circulatory motion, since our collective particle moves do not couple to the liquid velocity field, and are determined from the variations in the hight profile.

}

\section{Acknowledgements}

The authors would like to thank Tyler Shendruk and Uwe Thiele for valuable comments. MA gratefully acknowledges the KSA Ministry of Higher Education and University of Tabuk for financial support. {We are also grateful for constructive comments from two anonymous referees.}


\end{document}